\documentclass[10pt]{emulateapj}

\usepackage{rotating}

\newcommand{\Msun}{\ensuremath{{\rm M}_{\odot}}}

\shorttitle{Low-metallicity AGB evolution and nucleosynthesis}
\shortauthors{Fishlock et al.}

\begin{document}

\title{Evolution and nucleosynthesis of asymptotic giant branch stellar models of low metallicity}

\author{Cherie K. Fishlock}
\affil{Research School of Astronomy \& Astrophysics, Australian National University,
Canberra ACT 2611, Australia}
\email{cherie.fishlock@anu.edu.au}

\author{Amanda I. Karakas}
\affil{Research School of Astronomy \& Astrophysics, Australian National University,
Canberra ACT 2611, Australia}
\email{amanda.karakas@anu.edu.au}

\author{Maria Lugaro}
\affil{Monash Centre for Astrophysics, Monash University, Clayton VIC 3800, Australia}
\email{maria.lugaro@monash.edu}

\and

\author{David Yong}
\affil{Research School of Astronomy \& Astrophysics, Australian National University,
Canberra ACT 2611, Australia}
\email{david.yong@anu.edu.au}

\begin{abstract}
We present stellar evolutionary tracks and nucleosynthetic predictions for a grid of stellar models of low- and intermediate-mass asymptotic giant branch (AGB) stars at $Z=0.001$ ([Fe/H]$~=~-1.2$). The models cover an initial mass range from 1~$\Msun$ to 7~$\Msun$. Final surface abundances and stellar yields are calculated for all elements from hydrogen to bismuth as well as isotopes up to the iron group. We present the first study of neutron-capture nucleosynthesis in intermediate-mass AGB models, including a super-AGB model, of [Fe/H]~= $-1.2$. We examine in detail a low-mass AGB model of 2~$\Msun$ where the $^{13}$C($\alpha$,$n$)$^{16}$O reaction is the main source of neutrons. We also examine an intermediate-mass AGB model of  5~$\Msun$ where intershell temperatures are high enough to activate the $^{22}$Ne neutron source, which produces high neutron densities up to $\sim 10^{14}$~n~cm$^{-3}$. Hot bottom burning is activated in models with $M \geq 3$~$\Msun$. With the 3~$\Msun$ model we investigate the effect of varying the extent in mass of the region where protons are mixed from the envelope into the intershell at the deepest extent of each third dredge-up. We compare the results of the low-mass models to three post-AGB stars with a metallicity of [Fe/H] $\simeq -1.2$. The composition is a good match to the predicted neutron-capture abundances except for Pb and we confirm that the observed Pb abundances are lower than what is calculated by AGB models.
\end{abstract}


\section{Introduction}

Stars with an initial mass of between $\sim$0.8 and $\sim$8~$\Msun$, depending on initial metallicity, evolve through the asymptotic giant branch (AGB) phase. This is the last stage of nuclear burning for these stars \citep[for a review, see][]{2005ARA&A..43..435H,2006NuPhA.777..311S,Karakas:2014cn}. AGB stars are an observationally confirmed site for the \emph{slow} neutron-capture process \citep[the $s$-process, e.g.][]{2001ApJ...559.1117A}, which is responsible for the production of around half of the abundance of the heavy elements beyond Fe \citep{1998ApJ...497..388G}. AGB stars also produce a number of light elements such as lithium \citep[e.g.,][]{2010MNRAS.402L..72V}, carbon \citep[e.g.,][]{2009A&A...508.1359I}, fluorine \citep[e.g.,][]{2010ApJ...715L..94A,2012A&A...538A.117R}, and nitrogen \citep[e.g.,][]{Johnson:2007be}. Through nucleosynthesis and strong mass loss, AGB stars contribute to the chemical evolution of galaxies \citep{2007ApJ...659L..25M,2010A&A...522A..32R,2010A&A...523A..17L,2011MNRAS.414.3231K} as well as globular clusters \citep{2008A&A...479..805V,2009ApJ...699.2017M,2011A&A...532A...8M}. 

The stellar structure of an AGB star consists of an electron degenerate CO core surrounded by a He-burning shell and a H-burning shell. These shells are separated by the He-intershell which consists of approximately 75\% $^{4}$He, 22\% $^{12}$C, and 2\% $^{16}$O left over from partial He-burning. Surrounding the H-exhausted core (hereafter core) is a large convective envelope. Neutron-capture nucleosynthesis via the $s$-process takes place in the He-intershell where the abundance of $^{4}$He is high and  ($\alpha$,$n$) reactions can be efficiently activated releasing free neutrons that are then captured by the abundant $^{56}$Fe seed nuclei. The $s$-process terminates at Pb and Bi, the heaviest stable elements that can be produced with the low neutron densities that occur in AGB stars. For a review on $s$-process nucleosynthesis in AGB stars see \citet{1999ARA&A..37..239B}. 

During the thermally-pulsing AGB  phase, the star undergoes periodic thermal pulses (TPs) caused by instabilities in the thin He-burning shell. In order to liberate the energy that accumulates during He-burning, the He-burning shell drives a pulse-driven convective zone, which mixes ashes from the He-burning shell into the He-intershell. The energy released results in an expansion of the stellar layers above the CO core that effectively extinguishes the H-burning shell. This allows the convective envelope to move inwards in mass. If the convective envelope moves into the He-intershell, material enriched from partial He-burning and $s$-process nucleosynthesis is mixed to the surface. This mixing mechanism is known as the third dredge-up (TDU) and is one way of altering the surface composition of an AGB star.  Another important product that is mixed into the envelope is $^{12}$C from partial He-burning. Therefore the TDU is responsible for increasing the surface C/O ratio with the possibility of creating carbon-rich stars that have a C/O ratio greater than unity.

Nucleosynthesis in intermediate-mass AGB stars (M $\geq$ 3~$\Msun$ at $Z~=~0.001$) can also occur via proton captures at the base of the convective envelope. This mechanism is known as hot bottom burning (HBB). The temperature at the base of the convective envelope becomes sufficiently high which activates H-burning via the CNO cycle. If the temperature increases further, the Ne-Na chain and Mg-Al chain can also be activated \citep{1999A&A...347..572A}. One important consequence of HBB is the production of $^{14}$N at the expense of $^{12}$C and $^{16}$O, as well as decreasing the C/O ratio.

There are two main neutron source reactions in AGB stars: $^{13}$C($\alpha$,$n$)$^{16}$O and $^{22}$Ne($\alpha$,$n$)$^{25}$Mg. The $^{22}$Ne neutron source is efficiently activated at temperatures higher than approximately $300 \times 10^6$ K.  These temperatures are easily attained in the convective region that develops in the intershell during a TP for intermediate-mass stars. For low-mass stars, the $^{22}$Ne neutron source is only marginally activated and is ineffective in producing the neutrons required for substantial $s$-process nucleosynthesis. However, the $^{13}$C($\alpha$,$n$)$^{16}$O reaction is activated at temperatures as low as $90 \times 10^6$ K, which means it can be ignited in low-mass stars \citep{1995ApJ...440L..85S}. In canonical stellar models there is not enough $^{13}$C from the ashes of H-burning for it to be an efficient source of neutrons. In order to increase the abundance of $^{13}$C in the He-intershell, it is hypothesised that extra mixing of protons occurs at the deepest extent of the convective envelope during TDU. This is when  a sharp composition discontinuity forms where the H-rich envelope and He-intershell meet. Protons that have been mixed downwards are captured by $^{12}$C forming a ``pocket" of $^{13}$C  which usually burns in radiative conditions during the interpulse via the $^{13}$C($\alpha$,$n$)$^{16}$O reaction before the next TP. This releases free neutrons at densities of $\lesssim$~$10^8$~n cm$^{-3}$; much lower than the neutron densities reached by the $^{22}$Ne source of up to $\sim 10^{14}$~n~cm$^{-3}$. The total number of neutrons released (the neutron exposure), however, is higher for the $^{13}$C neutron source than the $^{22}$Ne neutron source because the neutron flux lasts for roughly $10^4$ years. For low-mass AGB stars, the $^{13}$C pocket is responsible for producing the bulk of the abundances of the $s$-process elements \citep[e.g.][]{2014ApJ...787...10B}.

The AGB phase terminates once the stellar envelope has been ejected as a result of strong mass loss with the CO core remaining as a white dwarf. The ejected material enriches the interstellar medium from which the next generation of stars form. 

The aim of this paper is to provide a self-consistent set of low- and intermediate-mass AGB models with [Fe/H]\footnote{[X/Y] = $\log_{10}(N_X/N_Y)_{\star} - \log_{10}(N_X/N_Y)_{\odot}$ where $N_X$ and $N_Y$ are the abundances of elements $X$ and $Y$.}~=~$-1.2$ appropriate for the study of dwarf spheroidal galaxies and globular clusters as well as direct comparison to post-AGB stars. The models can also provide input for synthetic and parametric studies \citep[e.g.][]{2004MNRAS.350..407I}. The models presented here are also applicable to investigating the pollution of Galactic halo stars by AGB stars and studies of galactic chemical evolution. We present evolution and nucleosynthesis results, including neutron-capture elements, for AGB models of 1 to 7~$\Msun$ for an initial metallicity of [Fe/H]~=~$-1.2$.  The models presented here cover the most extensive mass range of AGB stars at [Fe/H] = $-$1.2. In Section 2 we present the numerical details required for calculating the AGB stellar models. In Section 3 we present the stellar evolution results. The calculated models presented here provide the first detailed study of the TDU for an extended grid of AGB stars from 1 to 7~$\Msun$ at [Fe/H]~$=-1.2$.  In Section 4 we explore in more detail the evolution and nucleosynthesis results of a typical low-mass model (2~$\Msun$) and a typical intermediate-mass model (5~$\Msun$). In Section 5 we present the nucleosynthesis results including final surface abundances and stellar yields. In Section 6 we present the effect of varying the extent in mass of the region where protons are mixed from the envelope into the intershell for the 3~$\Msun$ model. In Section 7 we present a comparison between the low-mass model predictions and the observed abundances of three post-AGB stars. In Section 8 we discuss uncertainties in the stellar abundances and stellar yields as a result of assumptions in the input physics and we end with discussion and conclusions in Section 9.

\section{Evolutionary and nucleosynthesis codes}
\label{sec:code}

We calculate AGB stellar models for a range of initial masses from 1~$\Msun$ to 7~$\Msun$ with a metallicity of $Z~=~0.001$ ([Fe/H] = $-1.2$) and a helium abundance of $Y=0.25$. For the purposes of this study, we define the low-mass models to be those with an initial mass up to and including 3~$\Msun$, and the intermediate-mass models, 3.25~$\Msun$ and above. Each stellar model is evolved from the zero-age main sequence to near the end of the AGB phase when the majority of the convective envelope is lost by stellar winds. A two-step procedure is performed to calculate the structure and detailed nucleosynthesis for each stellar model. 

First, we use the Mt Stromlo Stellar Evolutionary code \citep[][and references therein]{2010ApJ...713..374K} to calculate the stellar evolutionary sequences.  The details of the procedure and evolution code are as described in \citet{2010ApJ...713..374K} except for the differences described below. For the low-mass models, we use the C and N enhanced \AE SOPUS low-temperature molecular opacity tables \citep{Marigo:2009gz} as used in \citet{2012ApJ...746...20K}. For the intermediate-mass models, we use updated \citet{Lodders:2003bf} scaled-solar \AE SOPUS low-temperature molecular opacity tables \citep{Marigo:2009gz}, which account for the depletion and enhancement of C and C/O. The opacity treatment utilised for the intermediate-mass models is described in detail in \citet{2014MNRAS.tmp....3F}. We use OPAL tables  \citep{1996ApJ...464..943I} updated to a \citet{Lodders:2003bf} scaled-solar abundance for consistency with the low-temperature opacity tables.

To model convective borders we follow the method described by \citet{1986ApJ...311..708L} and \citet{Frost:1996fu} which employs the Schwarzschild criterion but searches for a neutral border when $\nabla_{\rm ad}/\nabla_{\rm rad}$, the ratio of the adiabatic and radiative temperature gradients, is discontinuous such as during TDU. For convective regions we use the standard mixing length theory \citep{1958ZA.....46..108B} with a mixing length parameter of $\alpha= 1.86$. We use a solar global metallicity of $Z_{\odot} = 0.015$ with a scaled-solar initial composition from \citet{Asplund:2009eu} which has a protosolar metallicity of $0.0142$.  As with \citet{2010ApJ...713..374K}, mass loss prior to the AGB phase is included using the \citet{1975MSRSL...8..369R} formula with $\eta_R = 0.4$. Mass loss during the AGB phase is included using the \citet{Vassiliadis:1993jk} mass-loss prescription. 

Second, detailed nucleosynthesis calculations are performed using the stellar evolutionary sequences as input into a post-processing nucleosynthesis code \citep[see][and references therein for details]{Lugaro:2004en,2012ApJ...747....2L}. The nucleosynthesis code calculates nuclear reactions and mixing simultaneously to solve for the abundances. A post-processing code is necessary as the stellar evolutionary code only accounts for the major energy generating reactions involving H, $^3$He, $^4$He, $^{12}$C, $^{14}$N, and $^{16}$O. We assume the additional reactions included in the post-processing code produce negligible energy and do not affect the stellar structure \citep[see][]{2014MNRAS.437..195D}.

The updated nuclear network incorporated into the nucleosynthesis code is based on the JINA Reaclib\footnote{https://groups.nscl.msu.edu/jina/reaclib/db/} database as of May 2012 with the modifications as detailed in \citet{2014ApJ...780...95L}. The reaction rate of $^{13}$C($\alpha$,$n$)$^{16}$O is taken from \citet{2008PhRvC..78b5803H} while the reaction rates for $^{22}$Ne($\alpha$,$n$)$^{25}$Mg and $^{22}$Ne($\alpha$,$\gamma$)$^{26}$Mg are taken from \citet{2010NuPhA.841...31I}. The network, which considers 2336 reactions, includes 320 species from neutrons to polonium and comprises all the stable and unstable isotopes relevant for $s$-process nucleosynthesis (for example, we do not include the long-lived isotope $^{130}$Te, because it is not reached by the $s$-process). We further include two species for the unstable isotope $^{85}$Kr, the ground state $^{85}$Kr$^g$ and the short-lived metastable state $^{85}$Kr$^m$, due to their location at an $s$-process branching point. When determining surface abundances and yields, we assume that long-lived isotopes have decayed (e.g. $^{99}$Tc to $^{99}$Ru). 

For the low-mass models a partial mixing zone (PMZ) is included in the post-processing nucleosynthesis code. Protons are artificially added to the top layers of the He-intershell at the deepest extent of TDU where they are captured by $^{12}$C leading to the production of the $^{13}$C pocket \citep[see][]{2012ApJ...747....2L}. This produces the free neutrons required for $s$-process nucleosynthesis. For the low-mass models with an initial mass less than 2.75~$\Msun$, we choose the mass of the added PMZ  to be $2 \times 10^{-3}~\Msun$. For the 2.75 and 3~$\Msun$ models we choose a PMZ mass of $1 \times 10^{-3}~\Msun$ and $5 \times 10^{-4}~\Msun$, respectively. We choose a lower PMZ mass for the 2.75 and 3~$\Msun$ models because of the effect of a decreasing intershell mass with initial mass. We discuss the uncertainty related to the choice of the PMZ mass in Section~\ref{sec:pmz}. We set the mass to remain constant for every PMZ added during TDU. As the intershell mass reduces with each TP we take the neutron-capture abundances to be an upper limit.
 
\section{Stellar evolution results}

 In Table~\ref{tab:prop} we provide  a summary of the structural properties relevant for nucleosynthesis for each of the AGB models. We provide online tables for each model which include structural properties for each TP. Each online table  includes the pulse number, core mass, maximum mass of the intershell convection zone, duration of intershell convection, mass dredged into the envelope, the TDU efficiency, maximum temperature in the He-shell, maximum temperature at the base of the convective envelope during the previous interpulse period, maximum temperature in the H-shell during the previous interpulse period, interpulse period, total mass, maximum radiated luminosity during the interpulse period, maximum He-luminosity during a TP, maximum radius during the previous interpulse period, bolometric magnitude, and effective temperature at maximum radius.  Table~\ref{tab:dataset1} shows a portion of the table for each TP of the 1~$\Msun$ model and is published in its entirety in the electronic edition. 

\begin{table*}
\renewcommand{\arraystretch}{1.05}
 \begin{center}
  \caption{Evolutionary properties of the calculated $Z~=~0.001$ stellar models. 
 \label{tab:prop}}
  \vspace{1mm}
   \begin{tabular}{crrrrrrrrr}
   \tableline\tableline
$M_{\rm initial}$$^{\rm a}$	&	$M_{\rm final}$$^{\rm b}$	&	$M_{\rm core}$$^{\rm c}$	& $M_{\rm env}$$^{\rm d}$ & TPs$^{\rm e}$	&TDUs$^{\rm f}$  &	$\lambda_{\rm max}$$^{\rm g}$ &	$T_{\rm BCE}^{\rm max}$$^{\rm h}$&	$T_{\rm He}^{\rm max}$$^{\rm i}$	&	$M_{\rm TDU}$ $^{\rm j}$	\\
  ($\Msun$)  &        ($\Msun$) &        ($\Msun$)  &($\Msun$) &  &  &  & (10$^6$ K) &(10$^6$ K) & ($\Msun$)  \\
\tableline

1	&	0.678	&	0.667	&	0.011	&	17	&	2	&	0.08	&	1.4	&	284.1	&	0.002	   \\
1.25	&	0.669	&	0.649	&	0.020	&	14	&	8	&	0.16	&	2.2	&	271.8	&	0.009	   \\
1.5	&	0.657	&	0.646	&	0.011	&	14	&	10	&	0.37	&	6.7	&	275.4	&	0.026	   \\
2	&	0.668	&	0.661	&	0.007	&	17	&	14	&	0.73	&	4.2	&	294.2	&	0.095	   \\
2.25	&	0.839	&	0.673	&	0.166	&	17	&	16	&	0.82	&	5.6	&	305.4	&	0.132	   \\
2.5	&	0.948	&	0.709	&	0.239	&	17	&	16	&	0.92	&	9.2	&	318.5	&	0.138	   \\
2.75	&	1.057	&	0.746	&	0.312	&	18	&	18	&	0.97	&	15.8	&	320.8	&	0.138	   \\
3	&	1.189	&	0.792	&	0.397	&	22	&	20	&	1.00	&	28.3	&	332.5	&	0.124	   \\
3.25	&	1.403	&	0.843	&	0.561	&	23	&	22	&	1.00	&	48.9	&	350.5	&	0.093	   \\
3.5	&	1.176	&	0.857	&	0.319	&	27	&	27	&	0.99	&	58.5	&	361.3	&	0.104	   \\
4	&	1.726	&	0.883	&	0.843	&	68	&	68	&	1.02	&	82.9	&	361.3	&	0.231	   \\
4.5	&	1.659	&	0.908	&	0.750	&	79	&	78	&	0.97	&	87.6	&	356.6	&	0.210	   \\
5	&	1.740	&	0.938	&	0.802	&	94	&	93	&	0.95	&	92.5	&	361.2	&	0.194	   \\
5.5	&	1.962	&	0.972	&	0.990	&	100	&	99	&	0.93	&	98.1	&	363.0	&	0.151	   \\
6	&	1.725	&	1.015	&	0.709	&	108	&	105	&	0.92	&	104.8	&	376.5	&	0.107	   \\
7	&	2.062	&	1.145	&	0.917	&	135	&	132	&	0.86	&	125.0	&	392.4	&	0.034	   \\										
\tableline \tableline
  \end{tabular} 
\\
NB. -- a) the initial mass, b) the final mass, c) the final core mass, d) the final envelope mass, e) the number of TPs computed, f) the number of TDU episodes, g) the maximum efficiency of TDU, h) the maximum temperature reached at the base of the convective envelope, i) the maximum temperature reached at the base of the He-intershell, and j) the total amount of mass mixed into the envelope through TDU.
 \end{center}
\end{table*}

 The AGB phase is terminated when the stellar envelope is removed through mass loss.  The low-mass models, excluding the 1~$\Msun$ model, experience the superwind phase in the final few TPs during which the mass-loss rate reaches a plateau of approximately $10^{-5}$~$\Msun$~yr$^{-1}$.  The 1~$\Msun$ model loses the majority of its stellar envelope before it reaches the superwind phase. The intermediate-mass models experience the superwind phase well before most of the envelope has been lost. 

 For models with an initial mass up to (and including) 2~$\Msun$, we are able to evolve the envelope mass to less than 0.02~$\Msun$ which puts the model just beyond the tip of the AGB towards the post-AGB phase \citep{2001Ap&SS.275....1B}.  The models with $M \gtrsim 2$~$\Msun$ suffer from convergence problems towards the end of the AGB \citep[see][for more details]{2012A&A...542A...1L}. For the models between 2 and 4~$\Msun$ we are able to evolve the envelope mass to less than 0.6~$\Msun$. However, for the models between 4~$\Msun$ and 7~$\Msun$, we are able to evolve the envelope mass to less than 1~$\Msun$.

 Since some envelope mass still remains, it is possible that additional TDU episodes could occur which would further enrich the envelope prior to being ejected into the interstellar medium \citep[see][]{Karakas:2007gn}. If we assume that the mass lost during the final calculated interpulse period is taken as representative of the mass to be lost before the next possible TP, the models with $M \leq 3.5$~$\Msun$ cannot experience another TDU as there is not enough envelope mass left. The more massive models, however, retain sufficient mass to experience at least one more TDU episode. 
For example, the 6~$\Msun$~model has an envelope mass of 0.709~$\Msun$ remaining when calculations cease due to convergence issues. To estimate the number of remaining of TPs we assume the mass lost during the preceding TP is taken as representative of the mass to be lost in the following TPs (approximately $1.5 \times 10^{-1}$~$\Msun$). This leaves a minimum of an additional 4 TPs (possibly with TDU) that could take place.
 We remove the remaining envelope without taking into account the possibility for extra TDU(s). Therefore, the final surface abundance and yield predictions of the neutron-capture elements are a lower limit for the intermediate-mass models. Additionally, the termination of the intermediate-mass models occurs after HBB has ceased.
 
\begin{table}
\renewcommand{\arraystretch}{1.05}
 \begin{center}
  \caption{A tick (\checkmark) means the phenomenon occurred in each model, a cross ($\times$) if it did not.}
 \label{tab:results}
  \vspace{1mm}
   \begin{tabular}{cccccc}
   \tableline\tableline
 Mass  ($\Msun$)& Core He-flash  & FDU &  SDU & TDU & HBB  \\ 
\tableline
 1 & \checkmark& \checkmark & $\times$  & \checkmark & $\times$  \\
1.25 & \checkmark& \checkmark & $\times$ & \checkmark & $\times$   \\
 1.5& \checkmark &  \checkmark & $\times$  & \checkmark  & $\times$ \\
2 & $\times$& \checkmark  & $\times$   & \checkmark& $\times$  \\
2.25 & $\times$& \checkmark &  $\times$ & \checkmark   & $\times$  \\
2.5 & $\times$&  \checkmark &  $\times$ & \checkmark  & $\times$ \\
2.75 &  $\times$ &\checkmark & \checkmark  & \checkmark  & $\times$  \\ 
3  & $\times$  & \checkmark & \checkmark & \checkmark & \checkmark  \\
3.25 & $\times$ & \checkmark & \checkmark  & \checkmark & \checkmark \\
3.5  & $\times$  &  \checkmark  &  \checkmark  & \checkmark & \checkmark  \\
4 & $\times$  & $\times$ & \checkmark &  \checkmark &\checkmark \\ 
4.5  &  $\times$ & $\times$  &  \checkmark  & \checkmark & \checkmark  \\
5 & $\times$& $\times$& \checkmark  & \checkmark& \checkmark \\ 
5.5 & $\times$& $\times$& \checkmark  & \checkmark& \checkmark \\ 
6 & $\times$& $\times$& \checkmark  & \checkmark& \checkmark \\
7  & $\times$  & $\times$& \checkmark & \checkmark & \checkmark \\
\tableline \tableline
  \end{tabular} 
\\
 \end{center}
\end{table}

In Table~\ref{tab:results} we identify models which experience a core He-flash, the first dredge-up (FDU; after core H-burning at the base of the RGB), the second dredge-up (SDU; after core He-burning on the early AGB), TDU and/or HBB. All the models calculated with an initial mass less than 2~$\Msun$ develop an electron degenerate core and experience a core He-flash at the onset of core He-ignition. A core He-flash removes this degeneracy and the luminosity of the H-shell briefly reaches up to $10^9$ L$_{\odot}$. FDU is experienced in all models $\leq 3.5$~$\Msun$ (Table~\ref{tab:results}). The intermediate-mass models with $M \geq 4$~$\Msun$ do not experience FDU as core He-burning is ignited before the model reaches the first giant branch; these stars experience SDU as their first mixing episode. We find that SDU occurs in models with $M \geq 2.75~\Msun$. The 2.75 to 3.5~$\Msun$ models are the only models to experience both FDU and SDU.

All the stellar models experience TDU on the AGB and the efficiency of TDU is quantified by the parameter,

\begin{equation}
\lambda = \frac{\Delta M_{\rm dredge}}{\Delta M_{\rm c}},
\end{equation}

where $\Delta M_{\rm dredge}$ is the mass of the material mixed into the convective envelope by the TDU episode and $\Delta M_{\rm c}$ is the mass growth of the core due to H-burning during the preceding interpulse period. The 1~$\Msun$ model experiences the fewest number of TDUs with only six episodes which brings a total of 0.0016~$\Msun$ of enriched material to the stellar surface. The largest total amount of material that is mixed to the surface is 0.232~$\Msun$, which occurs for the 4~$\Msun$ model. Despite having more TDU episodes than the 4~$\Msun$ model, the 4.5, 5, 5.5, 6, and 7~$\Msun$ models dredge up a smaller amount of material. This is because the intershell region is not as massive and TDU is less efficient in these models compared to the 4~$\Msun$ model.  

\begin{figure}
\begin{center}
\includegraphics[width=\columnwidth]{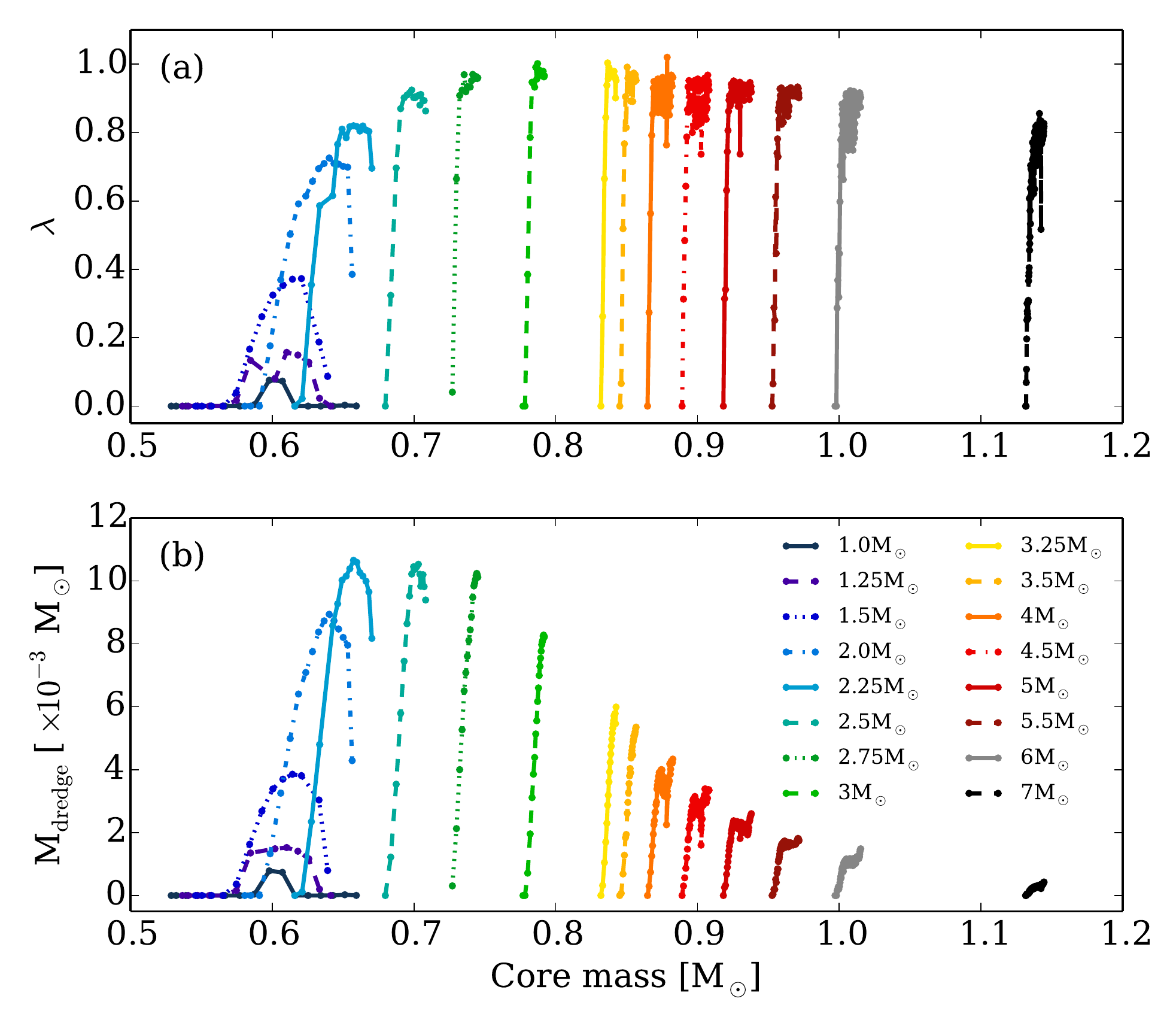} 
\caption{The evolution of  ($a$) the efficiency of TDU, $\lambda$ and ($b$) $M_{\rm dredge}$ with core mass for each model.}
\label{fig:f1}
\end{center}
\end{figure}

Figure~\ref{fig:f1}a illustrates the evolution of $\lambda$ with core mass for each model and the range of core masses produced by the models. The TDU efficiency gradually increases with increasing core mass. Overall, the efficiency of TDU increases with initial mass with the maximum $\lambda$ values occurring for the 3~and 3.25~$\Msun$ models. The overall efficiency then decreases for the 4.5, 5, 5.5, 6, and 7~$\Msun$ models. Figure~\ref{fig:f1}b reveals that the low-mass models, with the lowest $\lambda$ values, mix up more material per TDU as a result of a larger intershell mass compared to the intermediate-mass models. However, $M_{\rm dredge}$  does not correlate with $\lambda$ value. The increase in the core mass during the AGB phase is higher for the low-mass models as a result of a low $\lambda$. The mass of the core of the 1~$\Msun$ model increases by 0.14~$\Msun$ while the core mass of the 7~$\Msun$ model only increases by 0.014~$\Msun$, a factor of ten lower. This is a result of the high efficiency of TDU ($\lambda \approx 1$) and shorter interpulse periods in the intermediate-mass models leading to minimal core growth.

The maximum temperature reached at the base of the convective envelope increases with increasing initial mass reaching up to $125 \times 10^6$~K for the 7~$\Msun$ model (see Table~\ref{tab:prop}). While we find the lower initial stellar mass limit for HBB to be 3~$\Msun$ (Table~\ref{tab:results}) there is only mild activation of HBB for a few TPs in models less than 4~$\Msun$. The lower initial mass limit for efficient HBB with $T_{\rm bce} \gtrsim 60 \times 10^{6}$~K is 4~$\Msun$. The $Z~=~0.001$ models with an initial mass $\leq 2.5$~$\Msun$ of \citet{Ventura:2010bd}  do not experience HBB where they define the onset of HBB to be $T_{\rm bce} \gtrsim 60 \times 10^6$~K. 

The 6 and 7~$\Msun$ models experience hot TDU where HBB takes place during TDU as C and O is mixed the surface. The studies by \citet{2004A&A...421L..25G} and \citet{2004ApJS..155..651H} demonstrated that hot TDU can inhibit $s$-process nucleosynthesis. For these models, and the other intermediate-mass models, we do not include a PMZ.

The 7~$\Msun$ model is characterised as a super-AGB star as it experiences off-centre carbon ignition which produces an ONe core at the end of the AGB phase \citep{2007A&A...476..893S}. Super-AGB stars also experience high mass-loss rates with the 7~$\Msun$ model reaching a maximum rate of $1.3 \times 10^{-3}$~$\Msun$~yr$^{-1}$ after around 60 TPs. In comparison, the 6~$\Msun$ reaches a maximum mass-loss rate of $8.8 \times 10^{-4}$~$\Msun$~yr$^{-1}$ after around 80 TPs. 
The grid of super-AGB models calculated by \citet{2014MNRAS.441..582D} includes a 7~$\Msun$ model of $Z~=~0.001$. A comparison between these two models finds similar final core masses (1.14~$\Msun$ compared to 1.145~$\Msun$ for the model presented here), maximum temperature at the base of the convective envelope (120~MK compared to 125~MK), and total mass of material dredged up ($3.97 \times 10^{-2}$~$\Msun$ compared to $3.4 \times 10^{-2}$~$\Msun$). The 7~$\Msun$ model presented here experiences 135 TPs, whereas the \citet{2014MNRAS.441..582D} model experiences 126 TPs and is evolved to a smaller envelope mass.

\section{Details of a low- and intermediate-mass model}

\subsection{The 2~$\Msun$ model}
\label{sec:2msun}

The 2~$\Msun$ model was chosen as a representative case of low-mass AGB evolution at $Z~=~0.001$ as we are able to compare our results with the calculations of \citet{2009ApJ...696..797C,2011ApJS..197...17C}. Furthermore, we were able to evolve this model to a low envelope mass of 0.007~$\Msun$. In Table~\ref{tab:2msun_tp} we present, for each TP,  the total mass ($M_{\rm tot}$), the core mass ($M_{\rm core}$), the mass of material mixed to the surface due to TDU ($M_{\rm dredge}$), the efficiency of TDU ($\lambda$), the interpulse period ($\tau_{\rm ip}$), the maximum surface luminosity ($L_{\rm max}$), the effective temperature ($T_{\rm eff}$), the maximum radius ($R_{\rm max}$), and the surface C/O ratio (C/O).  The model experiences 17 TPs with 14 of these TPs followed by TDU.

\begin{figure}
\begin{center}
\includegraphics[width=\columnwidth]{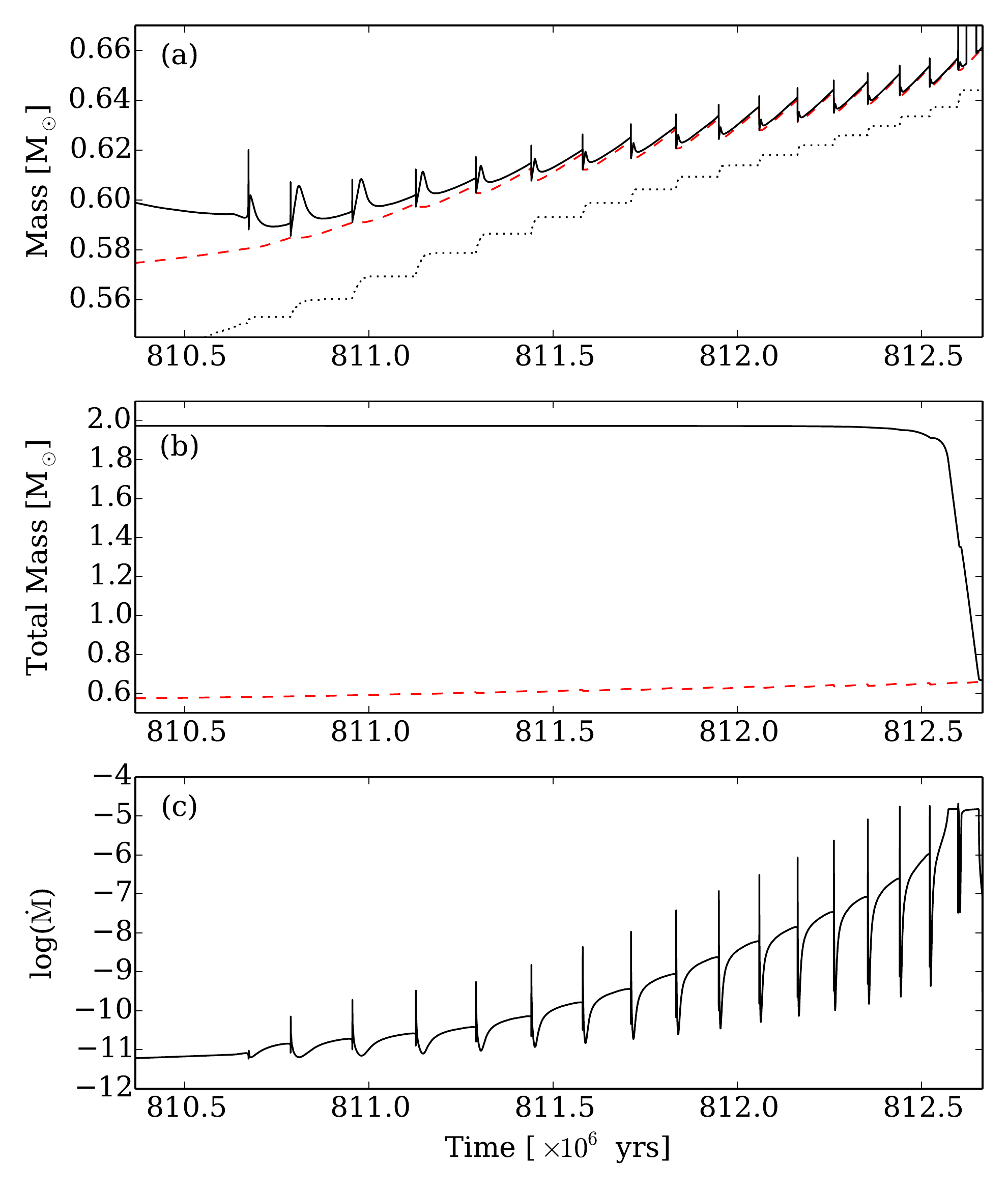} 
\caption{The temporal evolution from the start of the AGB phase for the 2~$\Msun$ model of ($a$) the inner edge of the convective envelope (solid line), the mass of the H-exhausted core (dashed line) and the mass of the He-exhausted core (dotted line), ($b$) total mass (solid line) and the mass of the H-exhausted core (dashed line) and ($c$) the mass-loss rate.}
\label{fig:f2}
\end{center}
\end{figure}

In Figure~\ref{fig:f2}a, we plot the temporal evolution of three different mass boundaries during the AGB phase: the inner edge of the convective envelope, the mass of the H-exhausted core, and the mass of the He-exhausted core. The mass of the He-exhausted core remains constant during the interpulse because the He-burning shell is mostly inactive. It is only during a TP that the He-exhausted core increases in mass. The mass of the H-exhausted core grows during the interpulse period when the H-burning shell is active. Following a TP the convective envelope moves inwards in mass and, if TDU occurs, H-rich material is mixed into the H-exhausted core thus reducing the mass of the core. 

\begin{table*}
 \begin{center}
  \caption{Properties of each TP for the 2~$\Msun$ model. }
 \label{tab:2msun_tp}
   \begin{tabular}{rccccccccr}
  \tableline\tableline
 TP  & $M_{\rm tot}$  & $M_{\rm core}$ & $M_{\rm dredge}$ & $\lambda$ &  $\log_{10}\tau_{\rm ip}$  & $\log_{10}L_{\rm max}$ & $\log_{10}T_{\rm eff}$ & $R_{\rm max}$ & C/O  \\ 
 &  ($\Msun$)  &        ($\Msun$) & ($\times 10^{-3} \Msun$)   & & (yr) & ($L_{\odot}$) & (K) &  ($R_{\odot}$) \\
\tableline
1	&	1.9737	&	0.581	&	0.00	&	0.00	&	$-$	&	3.36	&	3.64	&	85.41	&	0.29	 \\
2	&	1.9737	&	0.585	&	0.00	&	0.00	&	5.06	&	3.55	&	3.62	&	115.58	&	0.29	 \\
3	&	1.9737	&	0.591	&	0.00	&	0.00	&	5.22	&	3.62	&	3.62	&	128.64	&	0.29	 \\
4	&	1.9737	&	0.599	&	1.33	&	0.18	&	5.24	&	3.68	&	3.61	&	141.77	&	0.29	 \\
5	&	1.9737	&	0.606	&	3.25	&	0.37	&	5.21	&	3.74	&	3.61	&	156.02	&	0.83	 \\
6	&	1.9736	&	0.613	&	4.99	&	0.50	&	5.18	&	3.80	&	3.59	&	177.78	&	2.34	 \\
7	&	1.9736	&	0.619	&	6.40	&	0.59	&	5.15	&	3.85	&	3.58	&	202.51	&	4.43	 \\
8	&	1.9736	&	0.624	&	7.09	&	0.61	&	5.12	&	3.89	&	3.57	&	224.31	&	6.75	 \\
9	&	1.9735	&	0.628	&	7.75	&	0.66	&	5.09	&	3.92	&	3.55	&	246.33	&	8.78	 \\
10	&	1.9734	&	0.633	&	8.37	&	0.69	&	5.06	&	3.94	&	3.54	&	270.05	&	10.64	 \\
11	&	1.9730	&	0.637	&	8.73	&	0.71	&	5.04	&	3.96	&	3.53	&	291.06	&	12.41	 \\
12	&	1.9721	&	0.640	&	8.94	&	0.73	&	5.02	&	3.98	&	3.52	&	311.25	&	13.99	 \\
13	&	1.9702	&	0.644	&	8.73	&	0.71	&	4.99	&	4.00	&	3.52	&	330.66	&	15.54	 \\
14	&	1.9658	&	0.647	&	8.47	&	0.71	&	4.97	&	4.01	&	3.51	&	348.69	&	16.85	 \\
15	&	1.9543	&	0.650	&	8.20	&	0.70	&	4.94	&	4.03	&	3.50	&	367.14	&	18.03	 \\
16	&	1.9156	&	0.653	&	7.96	&	0.70	&	4.91	&	4.04	&	3.49	&	388.99	&	19.20	 \\
17	&	1.3980	&	0.656	&	4.29	&	0.39	&	4.89	&	4.05	&	3.45	&	518.40	&	20.20	 \\
\tableline \tableline
  \end{tabular} 
  \\ NB. -- The final mass is given in Table~\ref{tab:prop}.
 \end{center}
\end{table*}

The first TDU episode occurs after the fourth TP (once $\lambda$ is greater than zero). The dredge-up efficiency increases for each successive TP until it reaches a maximum value of $\lambda = 0.73$. TDU causes the C/O ratio to increase above unity by the sixth TP changing the envelope composition from oxygen rich to carbon rich (see Table~\ref{tab:2msun_tp}). The C/O ratio is approximately 20 by the last TP. A total amount of 0.0945~$\Msun$ of enriched material is mixed into the envelope through TDU (Table~\ref{tab:prop}), compared to 0.1313~$\Msun$ for the model calculated by \citet{2011ApJS..197...17C}. 

Appreciable envelope mass loss does not occur until the penultimate TP where the mass-loss rate increases to approximately $2 \times 10^{-5}~\Msun~{\rm yr}^{-1}$ during the superwind phase.  The superwind phase is where the majority of the envelope, around 1~$\Msun$, is lost. This is shown along with the temporal evolution of total mass and core mass  in Figures~\ref{fig:f2}b and~\ref{fig:f2}c. 

\begin{figure*}
\begin{center}
\includegraphics[width=2\columnwidth]{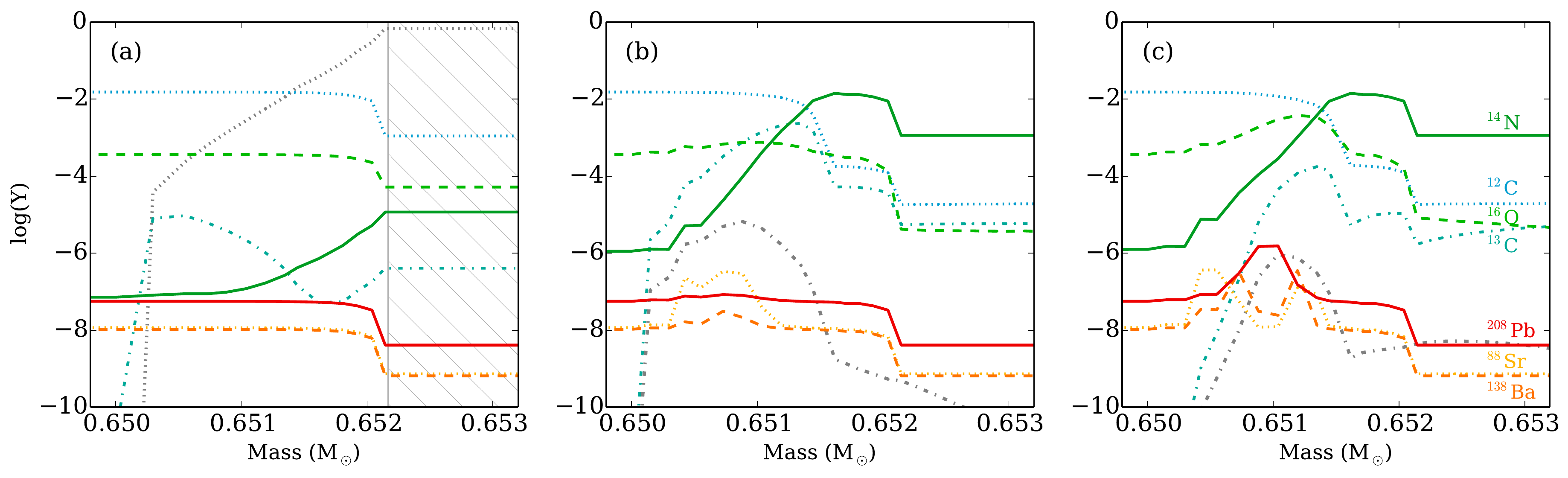} 
\caption{Composition profiles for three snapshots in the 2~$\Msun$ model after the last TDU. Abundances are given in units of $\log(Y)$, where $Y = X/A$ and $X$ is the mass fraction and $A$ is the atomic mass. Protons are shown by the grey dotted line. Neutrons are shown by the grey dash-dotted line and are offset in $\log(Y)$ by +15. The hatched region represents the convective envelope. Panel ($a$): the proton profile of the added PMZ just after the deepest extent of the last TDU. Panel ($b$): the $^{13}$C pocket has formed along with a $^{14}$N pocket with the $^{13}$C($\alpha$,$n$)$^{16}$O reaction producing a peak of neutrons. The neutron-capture elements are starting to be synthesised. Panel ($c$): Pb is created at the expense of Ba and the abundance of $^{13}$C  is now below that of $^{14}$N so no more neutrons can be produced.}
\label{fig:f3}
\end{center}
\end{figure*}

The composition profiles after the last TDU episode are presented in Figure~\ref{fig:f3} and focus on the region where the PMZ is added at the deepest extent of TDU. Key isotopes ($p$, $^{12}$C, $^{13}$C, $^{16}$O, $^{14}$N, $^{88}$Sr, $^{138}$Ba, and $^{208}$Pb) involved in $s$-process nucleosynthesis are presented. When the convective envelope reaches its most inward point in mass during TDU, a PMZ  of  $2 \times 10^{-3}$~$\Msun$ is added to the top of the He-intershell (illustrated in Figure~\ref{fig:f3}a). At the beginning of the interpulse, the $^{13}$C pocket forms along with a pocket of $^{14}$N. The mass of the $^{13}$C pocket is approximately $1 \times 10^{-3}$~$\Msun$. Later, the $^{13}$C($\alpha$,$n$)$^{16}$O reaction is activated increasing the neutron abundance.  In the regions where the $^{14}$N abundance is higher than the $^{13}$C abundance, no $s$-process nucleosynthesis can occur as  $^{14}$N acts as a neutron poison via the $^{14}$N($n$,$p$)$^{14}$C reaction (illustrated in Figure~\ref{fig:f3}b). Elements from the first peak, such as Sr, are produced first, followed by the second-peak elements such as Ba. Pb is then produced at the expense of these elements (illustrated in Figure~\ref{fig:f3}c). Eventually the abundance of $^{13}$C reduces to below that of $^{14}$N and $s$-process nucleosynthesis terminates. The enriched material is then mixed into the following TP and then to the stellar surface through the next TDU.

\begin{figure*}
\begin{center}
\includegraphics[width=\textwidth]{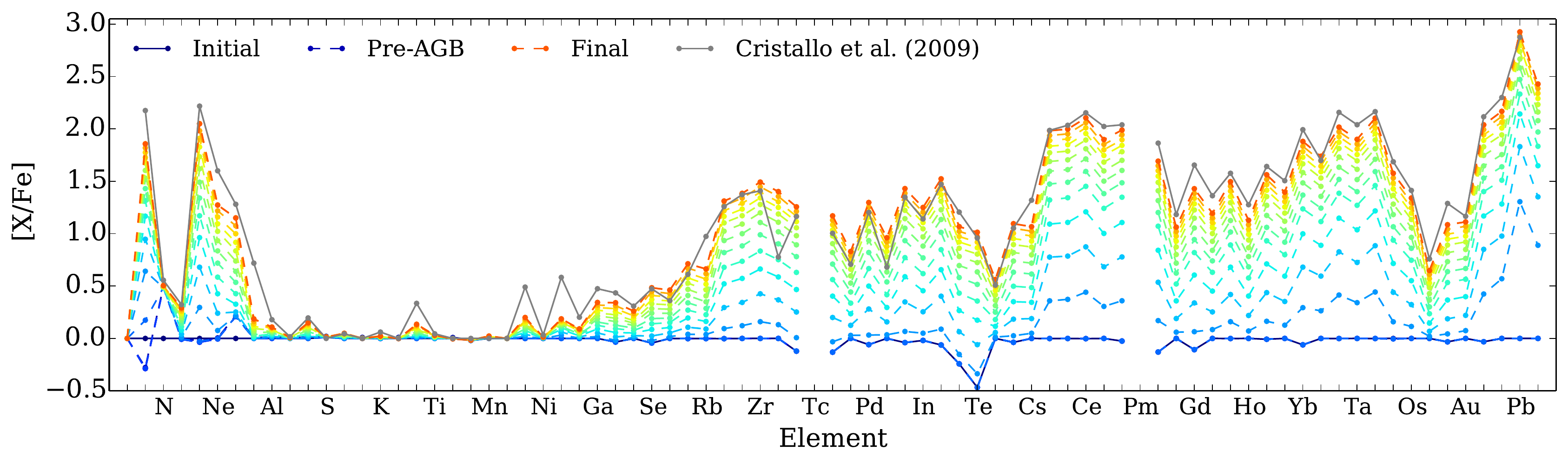} 
\caption{The surface abundance ratio for each of the elements after each TDU for the 2~$\Msun$ model. Each line connects the [X/Fe] abundance after each TDU. The initial and pre-AGB compositions are also shown. The initial composition shows that some elements have a value less than solar. This is a result of a few stable isotopes not being considered in the nuclear network as they are not accessible by the $s$-process (see Section~\ref{sec:code}). For comparison, the final surface abundance distribution for the 2~$\Msun$ model of \citet{2009ApJ...696..797C} is shown as a solid grey line. The elements are ordered by increasing atomic number. }
\label{fig:f4}
\end{center}
\end{figure*}

Figure~\ref{fig:f4} highlights the distribution of the surface abundance ratios relative to Fe for all elements from C to Bi. The `Initial' line is the initial composition on the main sequence. The `Pre-AGB' line is the pre-AGB composition as a result of FDU where the surface abundance of carbon, measured by [C/Fe], decreases by $0.28$~dex while [N/Fe] increases by 0.49~dex. The ratio of [Na/Fe] also increases by 0.22 dex. The remaining lines illustrate the surface abundances at the end of each TDU episode. The final abundances calculated by \citet{2009ApJ...696..797C} are also plotted.

Among the light elements, [C/Fe], [F/Fe], [Ne/Fe], and [Na/Fe] are enhanced by over 1~dex by the end of the AGB phase. The final [C/Fe] ratio is 1.88 as a result of TDU mixing up the products of partial He-burning. The [F/Fe] ratio increases from slightly below the solar value at the start of the AGB phase to 2.10, higher than the enhancement of [C/Fe]. The [O/Fe] value increases marginally  as a result of partial He-burning to 0.30 while the abundance of [N/Fe] only increases by 0.05~dex during the AGB phase. The final surface abundances of [Ne/Fe] and [Na/Fe] are enhanced to 1.3 and 1.2, respectively. The \citet{2009ApJ...696..797C} model has a higher enhancement in [Ne/Fe] and [Mg/Fe] with values up to 1.60 and 1.28, respectively. For the Fe-peak elements, there are minimal changes in the abundances. Both Co and Cu experience an enhancement of 0.2~dex while Sc increases by 0.14~dex and Zn by 0.1~dex. 

\begin{figure}
\begin{center}
\includegraphics[width=\columnwidth]{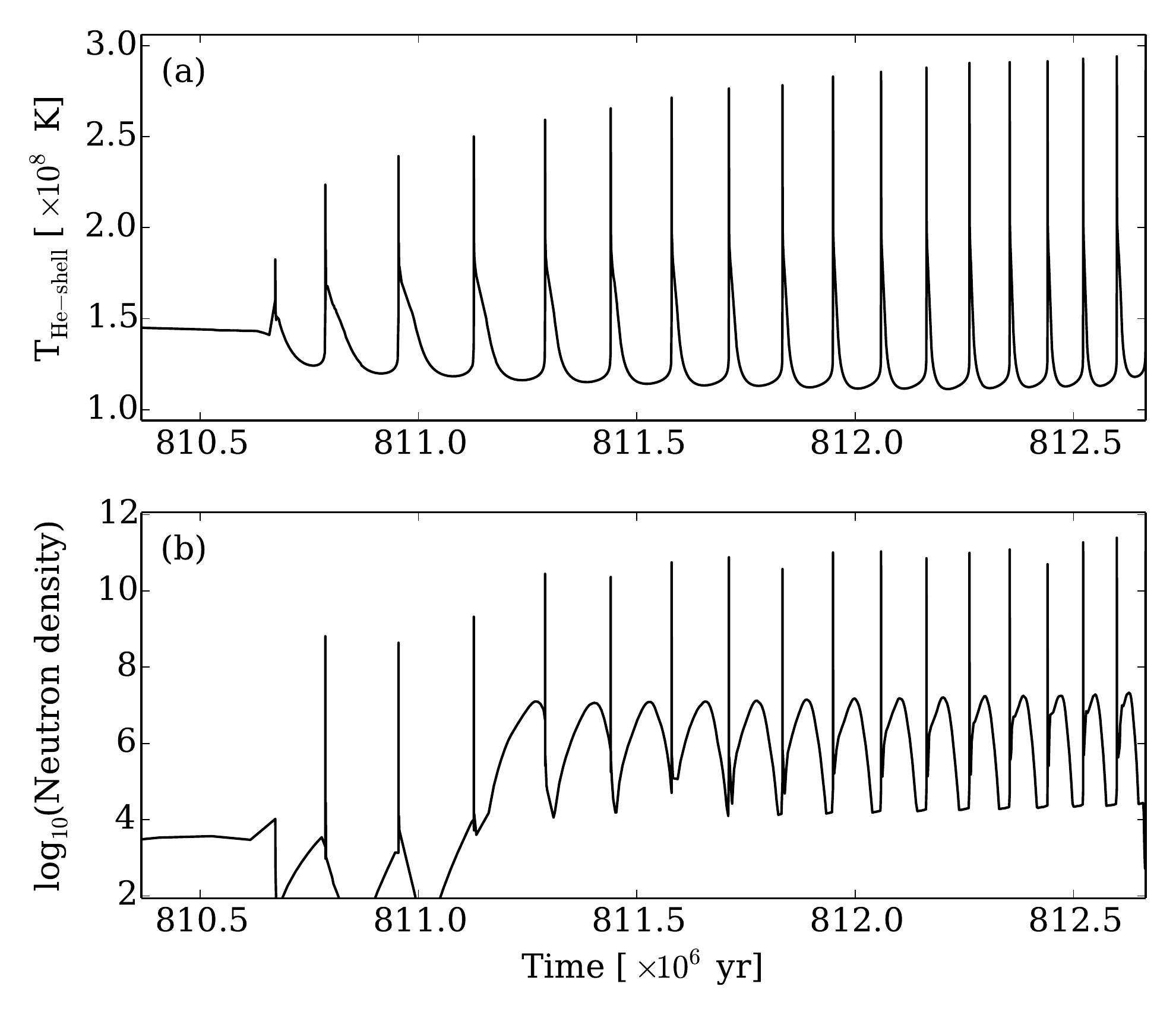} 
\caption{(a) The temperature at the top of the He-shell and (b) the maximum neutron density as a function of time from the start of the AGB phase for the 2~$\Msun$ model.}
\label{fig:f5}
\end{center}
\end{figure}

The surface abundance of the neutron-capture elements only increases after the second TDU episode (Figure~\ref{fig:f4}). The $^{13}$C pocket burns during the interpulse period once protons are added at the deepest extent of the first TDU. The newly synthesised neutron-capture elements are then mixed to the surface during the next TDU. The $s$-process abundance for each element asymptotically approaches its final value as each TDU brings more $s$-process enriched material to the surface. By the end of the AGB phase, the ratios of [Rb/Fe], [Zr/Fe], [Ba/Fe], and [Pb/Fe] are enhanced by 0.70, 1.53, 2.02, and 2.95~dex, respectively. These values follow a characteristic abundance distribution of neutron-capture elements for a low-metallicity low-mass AGB model where the production of Pb is favoured over the other neutron-capture elements \citep{2001ApJ...557..802B}. The model of \citet{2009ApJ...696..797C} produces 1.41, 1.99, and 2.87, respectively for [Zr/Fe], [Ba/Fe], and [Pb/Fe] and these values are comparable to those presented here despite a different treatment of the inner border of the convective envelope. The \citet{2009ApJ...696..797C} model has a noticeably higher enhancement of Rb as seen in Figure~\ref{fig:f4} as a result of neutron densities greater than $10^{12}$~n~cm$^{-3}$ occurring during a TP. The neutron densities in our model peak at less than $10^{11}$~n~cm$^{-3}$ during a TP (see Figure~\ref{fig:f5}) and have a minimal contribution to the abundance of Rb.

The surface abundance distribution for the neutron-capture elements exhibits three main peaks (around Sr, Ba, and Pb as seen in Figure~\ref{fig:f4}) corresponding to the isotopes with a magic number of neutrons ($N = 50, 82,126$). The average abundances of the neutron-capture elements at the first two of these points are called light $s$ ($ls$) and heavy $s$ ($hs$), respectively. The third peak is given as the abundance of [Pb/Fe].  As in \citet{2011ApJS..197...17C}, the [$ls$/Fe] abundance is given by,

\begin{equation}
{\rm [}ls{\rm/Fe] = ([Sr/Fe]+[Y/Fe]+[Zr/Fe])/3},
\end{equation}

and the [$hs$/Fe] abundance,

\begin{equation}
{\rm [}hs{\rm/Fe] = ([Ba/Fe]+[La/Fe]+[Nd/Fe]+[Sm/Fe])/4}.
\end{equation}

As the neutron exposure increases, the $ls$ elements are produced first, then the $hs$ elements, and finally Pb. For the 2~$\Msun$ model, the final surface abundance values of [$ls$/Fe], [$hs$/Fe], and [Pb/Fe] are 1.43, 1.95, and 2.95, respectively. Combinations of these ratios include [$hs$/$ls$] and [Pb/$hs$] and these $s$-process indicators are independent of the efficiency of TDU and the mass-loss rate for the low-mass models. For the model presented here the final [$hs$/$ls$] and [Pb/$hs$] values are 0.52 and 1.00.  These values are reached by the sixth TDU episode and remain constant until the end of the AGB phase.

The abundance distribution of the neutron-capture elements is predominately controlled by the neutron density, along with the neutron exposure. In Figure~\ref{fig:f5}, we plot, against time, the temperature of the He-burning shell and the maximum neutron density reached for each of the TPs. Notably, the first interpulse period with a $^{13}$C pocket has a neutron density around $10^7$~n~cm$^{-3}$. During this interpulse, not all of the $^{13}$C is burned radiatively and is later engulfed by the subsequent TP, resulting in convective $^{13}$C burning. This condition is described by  \citet{2012ApJ...747....2L} as Case 3. For the remaining $^{13}$C pockets, all the $^{13}$C is burned radiatively before the subsequent TP in accordance with Case 2 as described by \citet{2012ApJ...747....2L}. A peak in the neutron density occurs at each TP where there is a marginal activation of the $^{22}$Ne neutron source \citep{1998ApJ...497..388G}. The production of the neutron-capture elements from the $^{22}$Ne source is negligible compared to those produced from the $^{13}$C neutron source but elements produced via branching points such as Rb can be affected by this neutron flux.

\subsection{The 5~$\Msun$ model}

We examine the 5~$\Msun$ model as a representative case of an intermediate-mass AGB model. In contrast to a low-mass AGB model, the intermediate-mass models experience more efficient HBB and the $^{22}$Ne neutron source is  more efficiently activated. We do not add a PMZ for the reasons presented in detail by \citet{2013A&A...555L...3G}.

\begin{figure}
\begin{center}
\includegraphics[width=\columnwidth]{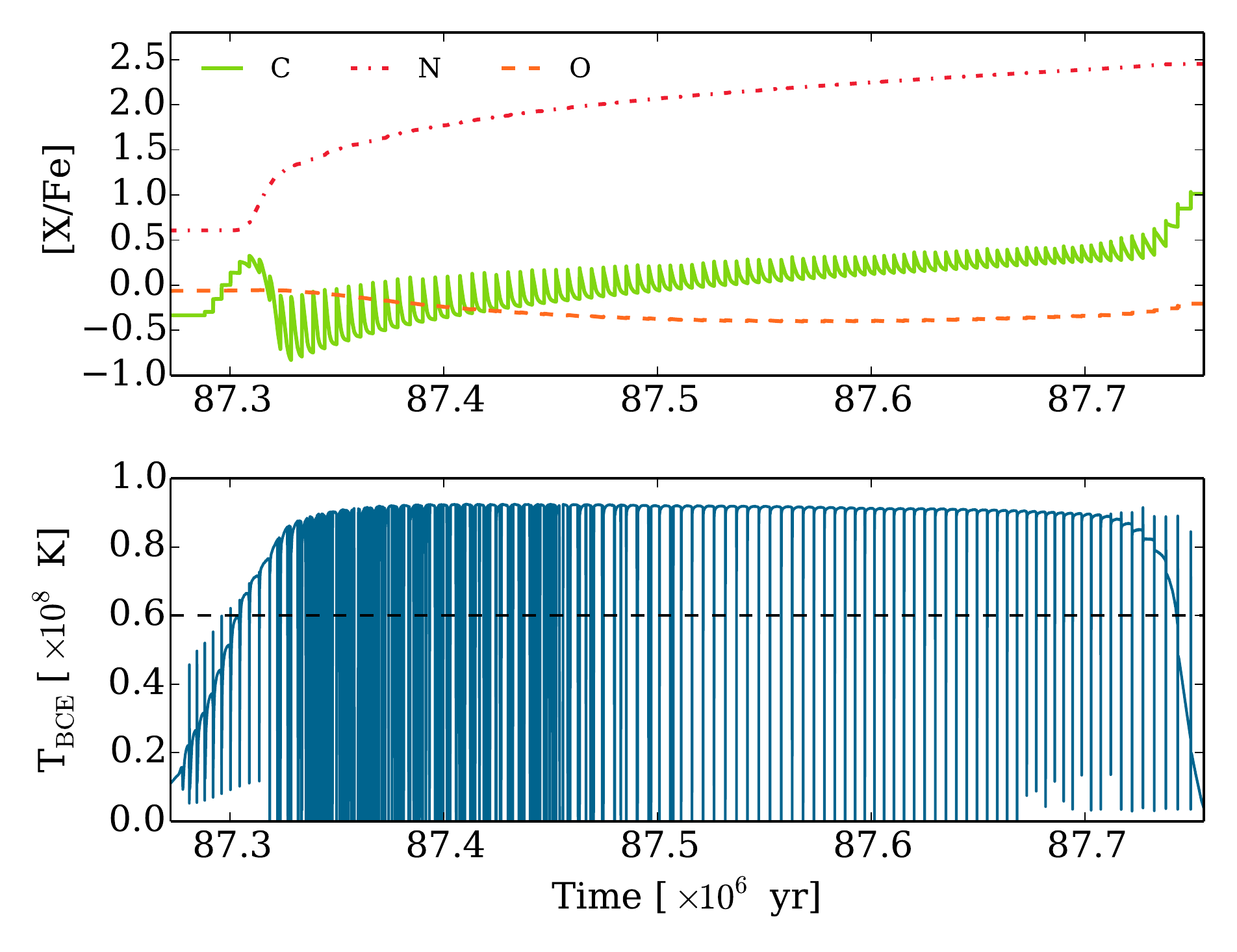} 
\caption{The evolution of [C/Fe], [N/Fe], and [O/Fe] (top) and the temperature at the base of the convective envelope (bottom)  with respect to time from the start of the AGB phase for the 5~$\Msun$ model. The dashed line illustrates the temperature above which HBB is efficiently activated.}
\label{fig:f6}
\end{center}
\end{figure}

 Figure~\ref{fig:f6} illustrates (a) the effect of HBB on the C, N, and O surface abundance relative to Fe and (b) the temperature at the base of the convective envelope. Initially, when the temperature at the base of the convective envelope has not reached the value required for CNO cycling, the ratio of [C/Fe] increases due to TDU while [N/Fe] remains constant. When temperatures at the base of the convective envelope reach approximately $60 \times 10^6$~K the CNO cycle is activated with C and O being converted to N. This destruction of C and O, along with the competing effect of TDU, causes the evolution of the C/O ratio to fluctuate about unity. From a certain point onwards, the model remains carbon rich. Eventually HBB is extinguished and [C/Fe] and [O/Fe] once again increases for the last few TDU episodes.

This fluctuating behaviour in the C/O ratio is also a feature in the 5~$\Msun$, $Z~=~0.001$ model of \citet{2013MNRAS.434..488M}. \citet{2013MNRAS.434..488M} also noted that the last few TPs do not experience HBB and the surface C/O ratio increases significantly to a final value of $\sim 10$. This value is comparable to the final C/O ratio of $\sim 9$ for the 5~$\Msun$ model presented here.  One difference, however, is that prior to HBB the  \citet{2013MNRAS.434..488M} model does not exhibit an increase in the C/O ratio in contrast to our model in which this ratio increases above unity before the onset of HBB (Figure~\ref{fig:f6}).

The efficiency of TDU in the 5~$\Msun$ model increases with each TP until it reaches a plateau of $\lambda \approx 0.95$  and the amount of material brought to the surface gradually increases until it reaches a value of approximately  $2.5 \times 10^{-3}$~$\Msun$. Despite the high efficiency of TDU, the amount of material mixed to the surface through each TDU is lower for the 5~$\Msun$ model compared to the low-mass model of 2~$\Msun$. This is a result of the He-intershell region  having approximately ten times less mass. However, the higher number of TDU episodes (93 compared to 14) means that the overall amount of the material being brought to the surface during the AGB phase is larger, 0.194~$\Msun$ compared to 0.095~$\Msun$ for the 2~$\Msun$ model (see Table~\ref{tab:prop}).

\begin{figure}
\begin{center}
\includegraphics[width=\columnwidth]{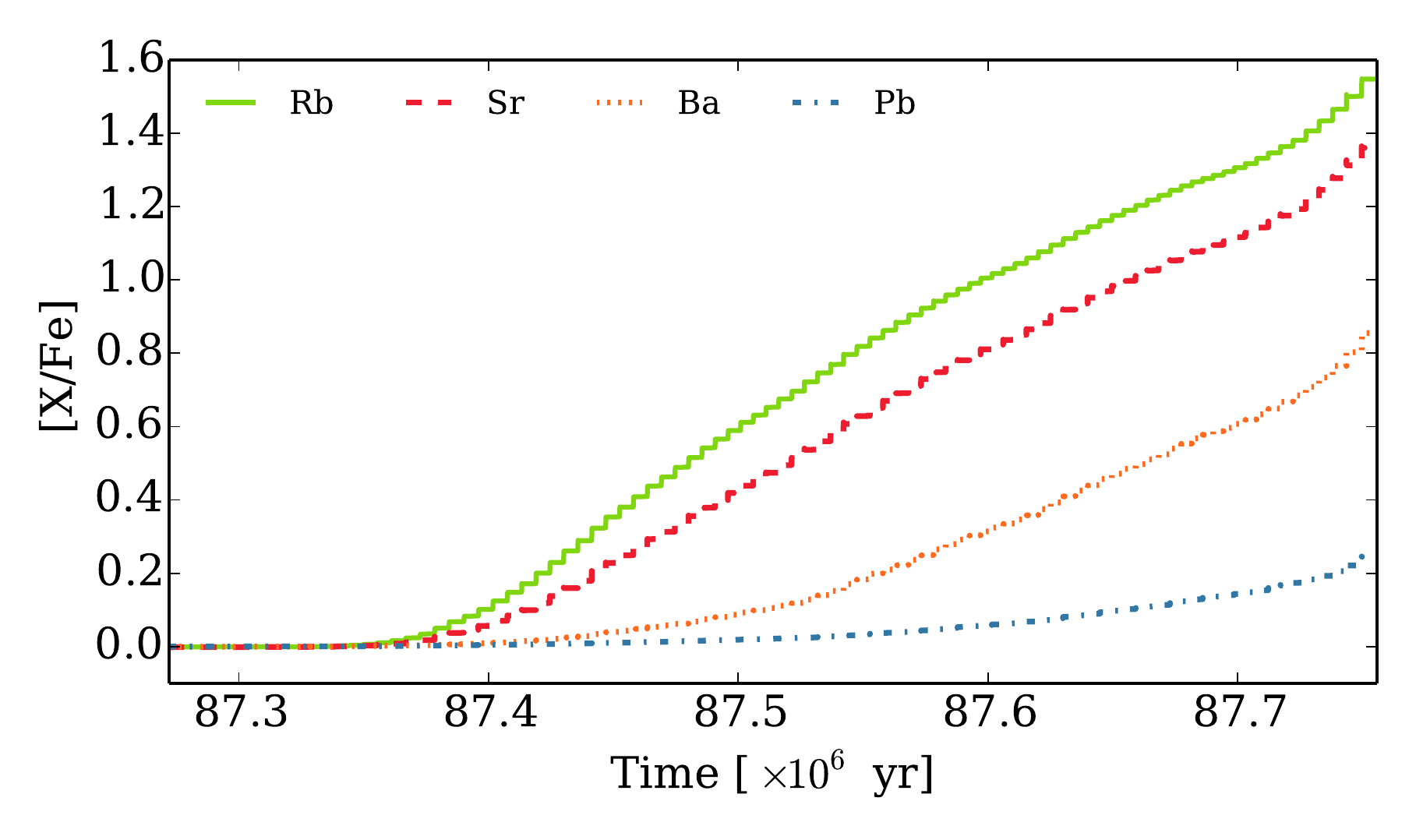} 
\caption{The surface abundance of four neutron-capture elements (Rb, Sr, Ba, and Pb) as a function of time from the start of the AGB phase for the 5~$\Msun$ model.}
\label{fig:f7}
\end{center}
\end{figure}

Figure~\ref{fig:f7}  presents the evolution of Rb, Sr, Ba, and Pb relative to Fe. These elements are representative of the three $s$-process peaks.  Rb exhibits the greatest enhancement and Pb the least and this situation is opposite to what the low-mass models display. The intermediate-mass models do not attain the neutron exposure required to produce elements such as Ba and Pb to the level produced by the low-mass models. However, the models are able to produce the high neutron densities required to bypass unstable isotopes (e.g., $^{85}$Kr) at branching points resulting in a higher abundance of Rb \citep[cf.][]{2012A&A...540A..44V}. The final [Rb/Fe] surface abundance is around 1.6 whereas [Pb/Fe] only increases by approximately 0.2~dex. 

\begin{figure}
\begin{center}
\includegraphics[width=\columnwidth]{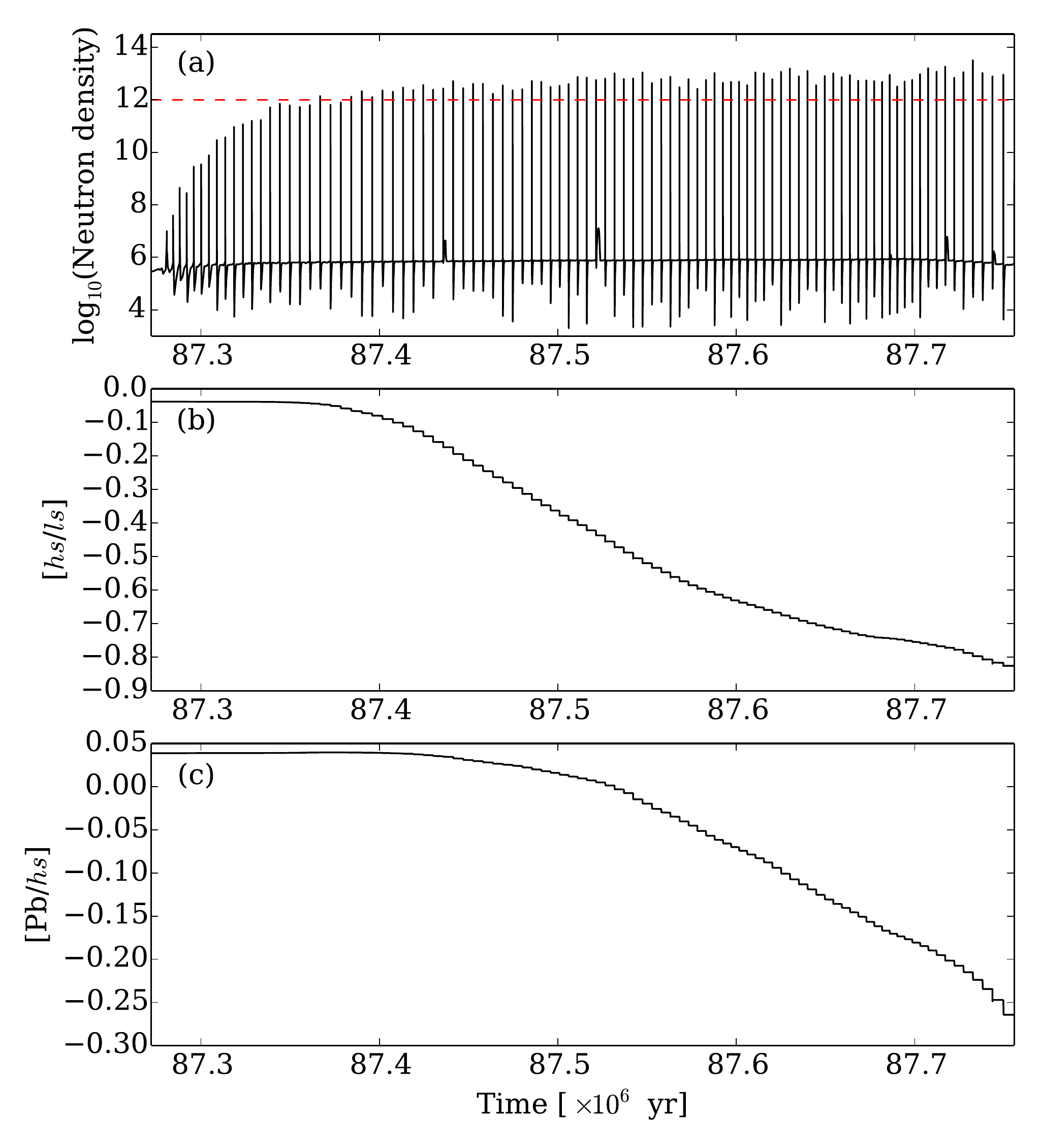} 
\caption{($a$) The maximum neutron density (the red dashed line shows the density where the $^{22}$Ne neutron source is efficient), ($b$) the surface [$hs$/$ls$] ratio, and ($c$) the surface [Pb/$hs$] ratio as a function of time for the 5~$\Msun$ model from the beginning of the AGB phase.}
\label{fig:f8}
\end{center}
\end{figure}

Figure~\ref{fig:f8}a illustrates the maximum neutron density reached for each TP as a function of time.  The peak neutron density is approximately $10^{13}$~n~cm$^{-3}$. The neutron density stays above $10^{12}$~n~cm$^{-3}$ for approximately 30 days  within each TP. As the temperature of the He-burning shell decreases, the neutron density also decreases. During the interpulse, the neutron density  remains constant just below $10^6$~n~cm$^{-3}$ due to the release of neutrons during radiative burning in the He-intershell. However, this neutron flux is not high enough to activate $s$-process nucleosynthesis.

The surface [$hs$/$ls$] and [Pb/$hs$] ratios from the beginning of the AGB phase  are presented in Figures~\ref{fig:f8}b and~\ref{fig:f8}c. The behaviour of the $s$-process indicators for the 5~$\Msun$ model (and all the intermediate-mass models) differs from the low-mass models due to the different neutron source. For the 5~$\Msun$ model, the ratios of [$ls$/Fe], [$hs$/Fe], [$hs$/$ls$], and [Pb/$hs$] remain constant during the AGB phase until the neutron density reaches above $10^{12}$~n~cm$^{-3}$. Once the neutron density exceeds this value the ratios of [$ls$/Fe] and [$hs$/Fe] increase while [$hs$/$ls$] and [Pb/$hs$] decrease. Unlike the low-mass models, the values of  [$hs$/$ls$] and [Pb/$hs$] never reach an equilibrium value. 

\subsection{Comparison with different metallicities}

We  briefly compare our 2 and 5~$\Msun$ models of [Fe/H] = $-1.2$ to the models of [Fe/H] = $-2.3$ presented in  \citet{2012ApJ...747....2L} and the models of [Fe/H] = $-1.7$ presented in \citet{2014ApJ...785...77S}. As the models of \citet{2014ApJ...785...77S} are $\alpha$ enhanced we only examine the neutron-capture elements (Ga to Bi) which are unaffected by the initial abundance of $\alpha$ elements.

\begin{figure*}
\begin{center}
\includegraphics[width=2\columnwidth]{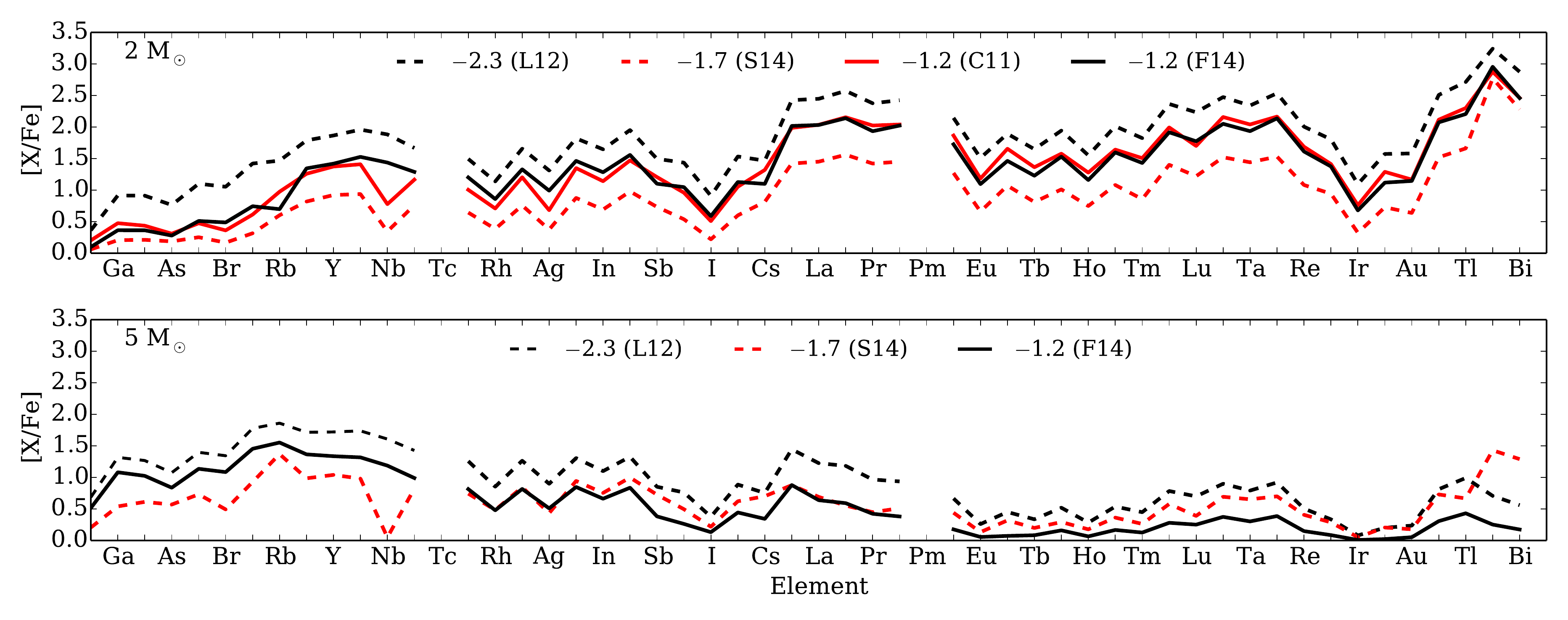} 
\caption{Final surface abundance ratios of neutron-capture elements for  three different [Fe/H] ratios for the 2~$\Msun$ models (top) and 5~$\Msun$ models (bottom). Legend is as follows: C11 \citep{2011ApJS..197...17C}, L12  \citep{2012ApJ...747....2L}, S14  \citep{2014ApJ...785...77S}, F14 (models presented here). The elements are ordered by increasing atomic number.}
\label{fig:f9}
\end{center}
\end{figure*}

Figure~\ref{fig:f9} illustrates the final neutron-capture surface abundances in [X/Fe] for each of the models. Both the models calculated here and in \citet{2012ApJ...747....2L} use the same evolutionary and post-processing codes, while the models of \citet{2011ApJS..197...17C} and \citet{2014ApJ...785...77S} use the FUNS code. The dip in abundance at Nb for the \citet{2011ApJS..197...17C} and \citet{2014ApJ...785...77S} models is due to the unstable isotope $^{93}$Zr ($\tau_{1/2} = 1.53 \times 10^6$ years) not being decayed to the stable isotope $^{93}$Nb. We include the 2~$\Msun$, $Z~=~0.001$ model from \citet{2011ApJS..197...17C} in Figure~\ref{fig:f9} for completeness.

Table~\ref{tab:compare_surf} presents a number of abundance ratios including the $s$-process indicators.  The models of \citet{2014ApJ...785...77S} have a [Fe/H] value in between the other models and we would expect them to show intermediate abundance values. However, the 2~$\Msun$ model appears to have essentially the same abundance pattern of the other models but with overall lower abundances. This is possibly the result of a lower amount of mass dredged up (0.075~$\Msun$) for their model, probably a consequence of the $\alpha$ enhancement employed in these models. The 5~$\Msun$ model of \citet{2014ApJ...785...77S} has a higher surface abundance of Pb compared to the 5~$\Msun$ model presented here and in \citet{2012ApJ...747....2L}.  This is due to the contribution of a small $^{13}$C pocket activated after each TDU  in the model of \citet{2014ApJ...785...77S}, which is not included in the other 5~$\Msun$ models.

\begin{table*}
   \def\arraystretch{1.2}
 \begin{center}
  \caption{Final surface abundances of the 2~and 5~$\Msun$ models for a number of neutron-capture elemental ratios for each metallicity.}
 \label{tab:compare_surf}
   \begin{tabular}{ccccccccc}
  \tableline\tableline
			& \multicolumn{2}{c}{L12} &	&  \multicolumn{2}{c}{S14} &	&  \multicolumn{2}{c}{F14} \\
			& \multicolumn{2}{c}{[Fe/H] = $-2.3$} &	&  \multicolumn{2}{c}{[Fe/H] = $-1.7$} &	&  \multicolumn{2}{c}{[Fe/H] = $-1.2$} \\
			\cline{2-3} \cline{5-6} \cline{8-9}
			
 	 		& 2~$\Msun$ 		 & 5~$\Msun$ &	& 2~$\Msun$ 	& 5~$\Msun$ &	& 2~$\Msun$ 	& 5~$\Msun$ \\
			\hline
			
$[$Rb/Fe$]$ & 1.47 & 1.86 & & 0.60 & 1.37 & & 0.70 & 1.55 \\
$[$Zr/Fe$]$ & 1.96 & 1.74&  & 0.94 & 0.98 & & 1.53 & 1.32 \\
$[$Rb/Zr$]$ & $-0.49$ & 0.12&  & $-$0.34 & 0.39 & & $-0.83$ & 0.24 \\
$[ls/$Fe$]$ & 1.87 & 1.72 & & 0.89 & 1.00 & & 1.43 & 1.34 \\
$[hs/$Fe$]$ & 2.36 & 1.07 & & 1.40 & 0.63 & & 1.95 & 0.52 \\
$[$Pb/Fe$]$ & 3.24 & 0.71&  & 2.77 & 1.43 & & 2.95 & 0.25 \\
$[hs/ls]$ & 0.49 & $-0.66$ & & 0.51 & $-0.37$ & & 0.52 & $-0.82$ \\
$[$Pb/$hs]$ & 0.88 & $-0.36$ &  & 1.37 & 0.80 & & 1.01 & $-0.27$ \\
\tableline
  \end{tabular} 
\\ NB. L12  \citep{2012ApJ...747....2L}, S14  \citep{2014ApJ...785...77S}, F14 (models presented here). 
 \end{center}
\end{table*}

\section{Abundance and stellar yield results}

 \begin{table*}
   \def\arraystretch{1.2}
 \begin{center}
  \caption{Net stellar yield results of selected light and neutron-capture elements for each low-mass model.}
 \label{tab:yields_lms}
   \begin{tabular}{crrrrrrrr}
  \tableline\tableline
El. & 1~$\Msun$ & 1.25~$\Msun$ & 1.25~$\Msun$ & 1.5~$\Msun$ & 2~$\Msun$ & 2.25~$\Msun$ & 2.5~$\Msun$ & 3~$\Msun$\\
\hline		
He &  8.56(-3) & 2.08(-2) & 3.16(-2) & 6.76(-2) & 8.48(-2) & 8.22(-2) & 8.25(-2) & 7.63(-2)   \\
C & 1.92(-4) & 1.57(-3) & 4.94(-3) & 1.65(-2) & 2.22(-2) & 2.34(-2) & 2.53(-2) & 2.44(-2)   \\
N & 1.95(-5) & 5.11(-5) & 8.09(-5) & 1.49(-4) & 1.72(-4) & 1.69(-4) & 1.85(-4) & 2.34(-4)   \\
O & 3.59(-6) & 6.82(-5) & 2.16(-4) & 5.34(-4) & 6.32(-4) & 6.34(-4) & 5.17(-4) & 4.71(-4)   \\
F & 5.63(-9) & 1.86(-7) & 7.27(-7) & 3.77(-6) & 6.40(-6) & 7.52(-6) & 6.56(-6) & 4.00(-6)   \\
Ne & 5.05(-6) & 1.41(-4) & 5.56(-4) & 2.45(-3) & 3.64(-3) & 4.03(-3) & 2.99(-3) & 2.01(-3)   \\
Na & 1.37(-7) & 1.49(-6) & 6.52(-6) & 3.65(-5) & 5.24(-5) & 4.29(-5) & 2.64(-5) & 2.09(-5)   \\
Mg & $-$2.17(-8) & 1.33(-6) & 6.49(-6) & 3.86(-5) & 6.92(-5) & 1.18(-4) & 1.44(-4) & 1.77(-4)   \\
Al & 6.74(-8) & 1.80(-7) & 3.75(-7) & 1.36(-6) & 2.68(-6) & 6.14(-6) & 7.46(-6) & 6.56(-6)   \\
Si & 1.22(-8) & 7.36(-8) & 1.82(-7) & 5.84(-7) & 1.16(-6) & 3.46(-6) & 5.99(-6) & 9.09(-6)   \\
Rb & 2.69(-11) & 5.76(-10) & 2.51(-9) & 6.89(-9) & 1.40(-8) & 3.41(-8) & 2.62(-8) & 2.41(-8)   \\
Sr & 5.51(-10) & 6.55(-9) & 2.68(-8) & 8.65(-8) & 1.25(-7) & 1.37(-7) & 8.33(-8) & 5.44(-8)   \\
Y & 1.15(-10) & 1.65(-9) & 6.29(-9) & 2.24(-8) & 3.36(-8) & 4.02(-8) & 2.52(-8) & 1.55(-8)   \\
Zr & 2.33(-10) & 4.64(-9) & 1.65(-8) & 6.23(-8) & 9.57(-8) & 1.19(-7) & 8.14(-8) & 5.41(-8)   \\
Ba & 1.35(-10) & 9.32(-9) & 3.09(-8) & 1.32(-7) & 2.02(-7) & 2.23(-7) & 1.34(-7) & 8.69(-8)   \\
La & 1.44(-11) & 9.23(-10) & 3.14(-9) & 1.35(-8) & 2.07(-8) & 2.28(-8) & 1.30(-8) & 7.81(-9)   \\
Ce & 5.85(-11) & 2.87(-9) & 1.04(-8) & 4.49(-8) & 7.03(-8) & 8.00(-8) & 4.80(-8) & 2.94(-8)   \\
Pb & 4.69(-9) & 1.19(-7) & 4.14(-7) & 1.27(-6) & 1.65(-6) & 1.64(-6) & 1.27(-6) & 9.11(-7)   \\
\tableline
  \end{tabular} 
\\ NB. Yields are in solar masses and are expressed in the form $n(m) = n \times 10^m$.
 \end{center}
\end{table*}

In this section we present final surface abundances and elemental stellar yields for each of the calculated models.  We calculate the net stellar yield $M_i$ (in M$_{\odot}$) to be,

\begin{equation}
M_i = \int_0^{\tau} \left [ X(i) - X_0(i) \right ] \frac{dM}{dt} dt,
\label{eq:yield}
\end{equation}

where $dM/dt$ is the current mass-loss rate in~$\Msun$ yr$^{-1}$, $X(i)$ and $X_0(i)$ are the current and initial mass fraction of species $i$, and $\tau$ is the total lifetime of the stellar model \citep{Karakas:2010et}. For a negative net yield, the species is destroyed whereas a positive net yield indicates that the species is produced. Tables~\ref{tab:yields_lms} and~\ref{tab:yields_ims} present net yields of select elements for each model.  We have made available online tables presenting yields for isotopes up to the Fe group and all the elements.   We provide for each model: the species $i$, the atomic number, the net stellar yield as defined in Equation~\ref{eq:yield}, the amount of the species $i$ in the wind lost from the star which is always positive, and the total mass expelled during the stellar lifetime multiplied by the initial mass fraction, $M_0(i)$. We also include the average mass fraction of $i$ in the wind $\langle X(i)\rangle$, the initial mass fraction $X_0(i)$, and the production factor $f$ defined as $\log_{10} [\langle X(i)\rangle/X_0(i)]$. Tables~\ref{tab:dataset3} and~\ref{tab:dataset5} show a portion of the yields table for the isotopes and elements and is published in its entirety in the electronic edition.  In addition, Tables~\ref{tab:dataset2} and~\ref{tab:dataset4} show a portion of the final surface abundances table for the isotopes and elements and is published in its entirety in the electronic edition.

\subsection{The light elements}

\begin{figure}
\begin{center}
\includegraphics[width=\columnwidth]{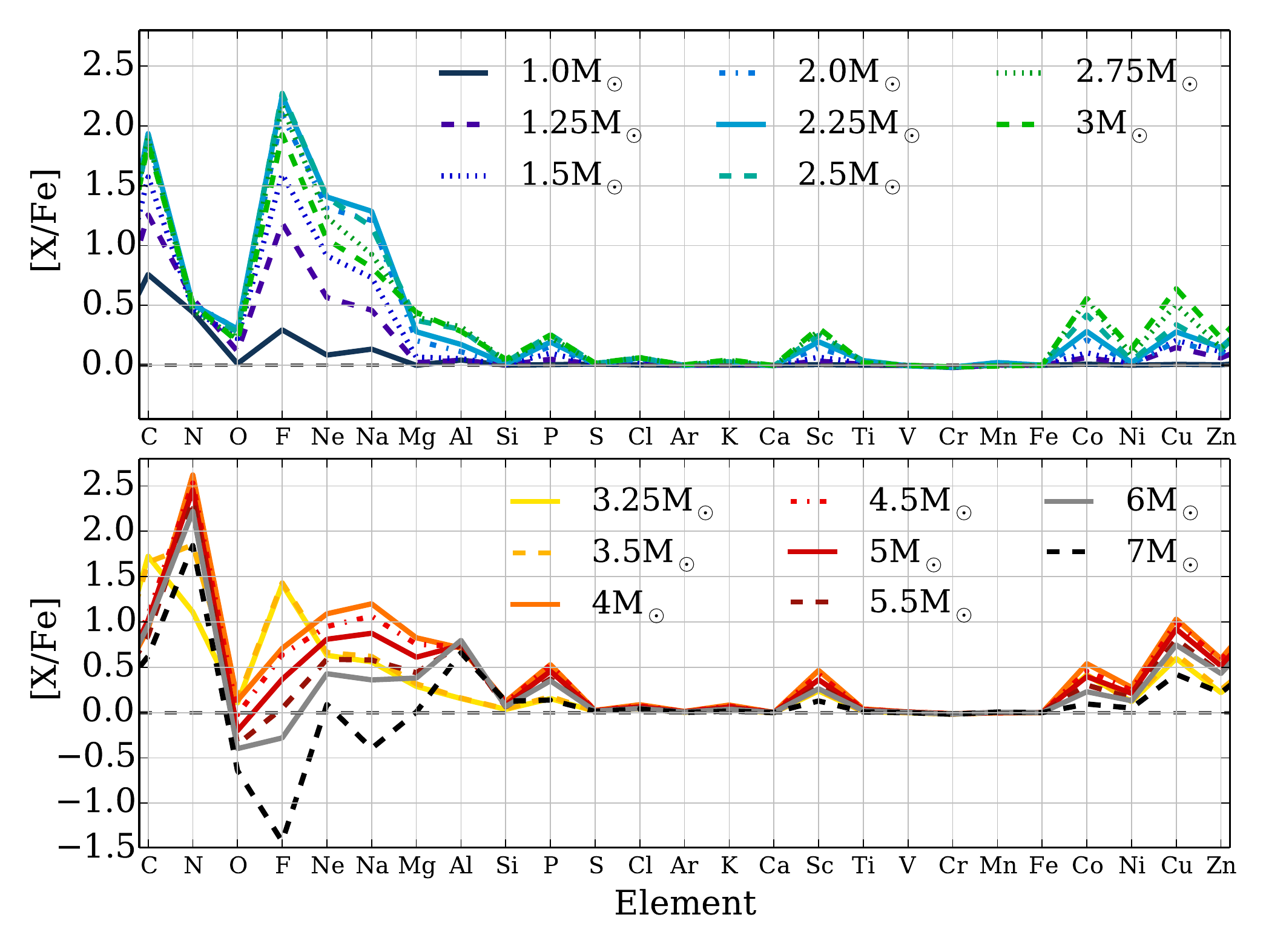} 
\caption{Final surface abundances relative to Fe for each of the models for the light elements from C to Zn. The elements are ordered by increasing atomic number.}
\label{fig:f10}
\end{center}
\end{figure}

In this section we present final surface abundances and yields for the light elements up to the Fe group. We discuss He, C, N, O, F, Ne, Na, Mg, and Al in detail. These elements are well known to be produced or destroyed in AGB stars \citep{1999ARA&A..37..239B}. In Figure~\ref{fig:f10} we present the final surface abundances of select light elements  (in [X/Fe]) for each of the models. Table~\ref{tab:lightelements} presents the final surface abundances for the $^4$He mass fraction, C/O ratio, $^{12}$C/$^{13}$C ratio, and  [X/Fe] for the selected light elements.

 \begin{table*}
   \def\arraystretch{1.2}
 \begin{center}
  \caption{Net stellar yield results of selected light and neutron-capture elements for each intermediate-mass model.}
 \label{tab:yields_ims}
   \begin{tabular}{crrrrrrrr}
  \tableline\tableline
El. & 3.25~$\Msun$ & 3.5~$\Msun$ & 4~$\Msun$ & 4.5~$\Msun$ & 5~$\Msun$ & 5.5~$\Msun$ & 6~$\Msun$ & 7~$\Msun$\\
\hline
He & 6.89(-2) & 1.01(-1) & 2.66(-1) & 3.57(-1) & 4.39(-1) & 5.01(-1) & 5.48(-1) & 6.06(-1)   \\
C & 1.95(-2) & 1.56(-2) & 2.39(-3) & 3.04(-3) & 2.63(-3) & 1.80(-3) & 1.74(-3) & 2.56(-4)   \\
N & 1.24(-3) & 8.53(-3) & 5.94(-2) & 5.61(-2) & 5.22(-2) & 4.33(-2) & 3.29(-2) & 1.26(-2)   \\
O & 2.19(-4) & 2.99(-4) & 3.09(-4) & $-$2.62(-4) & $-$8.28(-4) & $-$1.24(-3) & $-$1.61(-3) & $-$2.20(-3)   \\
F & 1.27(-6) & 1.28(-6) & 1.45(-7) & 7.97(-8) & $-$4.54(-9) & $-$6.43(-8) & $-$1.02(-7) & $-$1.41(-7)   \\
Ne & 6.58(-4) & 6.77(-4) & 2.93(-3) & 2.21(-3) & 1.64(-3) & 9.32(-4) & 4.84(-4) & 5.81(-5)   \\
Na & 1.04(-5) & 1.26(-5) & 7.57(-5) & 5.94(-5) & 3.86(-5) & 1.64(-5) & 3.65(-6) & $-$8.63(-6)   \\
Mg & 9.45(-5) & 1.02(-4) & 7.53(-4) & 6.56(-4) & 4.47(-4) & 2.75(-4) & 1.39(-4) & $-$6.84(-5)   \\
Al & 3.14(-6) & 3.15(-6) & 4.01(-5) & 4.88(-5) & 5.86(-5) & 6.89(-5) & 8.55(-5) & 7.00(-5)   \\
Si & 6.73(-6) & 7.63(-6) & 3.37(-5) & 3.19(-5) & 2.68(-5) & 2.22(-5) & 2.57(-5) & 7.73(-5)   \\
Rb & 1.48(-8) & 1.72(-8) & 1.53(-7) & 1.64(-7) & 1.38(-7) & 1.12(-7) & 8.50(-8) & 2.41(-8)   \\
Sr & 1.72(-8) & 2.05(-8) & 2.20(-7) & 2.38(-7) & 2.04(-7) & 1.67(-7) & 1.22(-7) & 3.06(-8)   \\
Y & 2.96(-9) & 3.56(-9) & 4.40(-8) & 4.85(-8) & 4.05(-8) & 3.38(-8) & 2.40(-8) & 5.79(-9)   \\
Zr & 5.23(-9) & 6.28(-9) & 8.99(-8) & 1.01(-7) & 8.27(-8) & 6.93(-8) & 4.78(-8) & 1.09(-8)   \\
Ba & 7.75(-10) & 9.17(-10) & 2.01(-8) & 2.28(-8) & 1.78(-8) & 1.50(-8) & 9.15(-9) & 1.77(-9)   \\
La & 3.26(-11) & 4.03(-11) & 1.06(-9) & 1.17(-9) & 8.93(-10) & 7.55(-10) & 4.46(-10) & 8.11(-11)   \\
Ce & 7.83(-11) & 9.41(-11) & 2.41(-9) & 2.71(-9) & 2.01(-9) & 1.68(-9) & 9.64(-10) & 1.75(-10)   \\
Pb & 1.60(-10) & 1.79(-10) & 2.51(-9) & 2.90(-9) & 2.27(-9) & 1.82(-9) & 9.65(-10) & 1.72(-10)   \\
\tableline
  \end{tabular} 
\\ NB. Yields are in solar masses and are expressed in the form $n(m) = n \times 10^m$.
 \end{center}
\end{table*}

  \begin{table*}
   \def\arraystretch{1.2}
 \begin{center}
  \caption{Final surface $^4$He mass fraction, C/O ratio, $^{12}$C/$^{13}$C ratio, and [X/Fe] ratios for selected light elements. }
 \label{tab:lightelements}
   \begin{tabular}{crccrrrrrrrr}
  \tableline\tableline
Mass & $^{4}$He & C/O & $^{12}$C/$^{13}$C & [C/Fe] & [N/Fe] & [O/Fe] & [F/Fe] & [Ne/Fe] & [Na/Fe] & [Mg/Fe] & [Al/Fe] \\
\hline
1.00 & 0.28 & 3.07 & 160  & 0.76 & 0.44 & 0.01 & 0.30 & 0.09 & 0.13 & 0.00 & 0.04   \\
1.25 & 0.29 & 7.67 & 575 & 1.26 & 0.45 & 0.12 & 1.19 & 0.57 & 0.40 & 0.02 & 0.04   \\
1.50 & 0.29 & 12.47 & 1232 & 1.58 & 0.47 & 0.22 & 1.60 & 0.91 & 0.73 & 0.07 & 0.05   \\
2.00 & 0.30 & 21.04 & 2687 & 1.88 & 0.51 & 0.30 & 2.10 & 1.31 & 1.21 & 0.21 & 0.11   \\
2.25 & 0.31 & 23.80 & 3129 & 1.94 & 0.51 & 0.30 & 2.25 & 1.41 & 1.29 & 0.28 & 0.17   \\
2.50 & 0.30 & 23.47 & 2752 & 1.91 & 0.46 & 0.28 & 2.27 & 1.40 & 1.16 & 0.37 & 0.30   \\
2.75 & 0.29 & 26.02 & 2355 & 1.90 & 0.46 & 0.22 & 2.17 & 1.24 & 0.93 & 0.40 & 0.32   \\
3.00 & 0.29 & 24.82 & 89 & 1.85 & 0.51 & 0.19 & 1.92 & 1.05 & 0.81 & 0.44 & 0.28   \\
3.25 & 0.28 & 22.93 & 11 & 1.72 & 1.11 & 0.10 & 1.41 & 0.63 & 0.56 & 0.29 & 0.15   \\
3.50 & 0.29 & 18.32 & 6.27 & 1.66 & 1.85 & 0.13 & 1.43 & 0.66 & 0.62 & 0.31 & 0.16   \\
4.00 & 0.34 & 3.38 & 5.97 & 0.91 & 2.62 & 0.12 & 0.70 & 1.08 & 1.19 & 0.82 & 0.70   \\
4.50 & 0.35 & 6.91 & 7.99 & 1.07 & 2.53 & $-$0.03 & 0.63 & 0.94 & 1.06 & 0.75 & 0.72   \\
5.00 & 0.36 & 9.07 & 7.56 & 1.01 & 2.45 & $-$0.20 & 0.36 & 0.80 & 0.87 & 0.61 & 0.74   \\
5.50 & 0.36 & 8.69 & 6.11 & 0.84 & 2.34 & $-$0.35 & 0.04 & 0.58 & 0.58 & 0.44 & 0.75   \\
6.00 & 0.36 & 12.58 & 8.53 & 0.96 & 2.22 & $-$0.40 & $-$0.28 & 0.43 & 0.36 & 0.38 & 0.79   \\
7.00 & 0.36 & 10.09 & 6.65 & 0.62 & 1.84 & $-$0.64 & $-$1.42 & 0.09 & $-$0.40 & 0.00 & 0.66   \\
\tableline
  \end{tabular} 
\\ 
 \end{center}
\end{table*}

\subsubsection{He, C, N, O, and F}

As presented in Table~\ref{tab:lightelements}, the final $^{4}$He surface abundance for the low-mass models reaches a peak value of 0.31 for the 2.25~$\Msun$ model. This is a result of the 2.25~$\Msun$ model experiencing the deepest extent of FDU and efficient TDU. The 2.25~$\Msun$ model also has the highest He yield of the low-mass models with a value of $8.48 \times 10^{-2}$~$\Msun$ (see Table~\ref{tab:yields_lms}). The $^4$He abundance reaches a maximum of 0.36 for the 5, 5.5, 6, and 7~$\Msun$ models and the yield increases with increasing initial mass, with a maximum He yield of $6.06 \times 10^{-1}$~$\Msun$ for the 7~$\Msun$ model.

The highest final surface abundance of [C/Fe] occurs for the 2.25~$\Msun$ model with a value of 1.94. The 7~$\Msun$ has the lowest final abundance due to very efficient HBB and a low $M_{\rm TDU}$ value (see Table~\ref{tab:prop}). The C yield increases with increasing initial mass for the models up to 2.75~$\Msun$. This increase in the yield follows the increase in the value of $M_{\rm TDU}$ where more C is mixed to the surface. The highest yield of C occurs for the 2.75~$\Msun$ model as it has the maximum $M_{\rm TDU}$ for the low-mass models. For the intermediate-mass models, the yield on the whole decreases with the 7~$\Msun$ model having the lowest yield of $2.56 \times 10^{-4}$~$\Msun$. 

In Table~\ref{tab:lightelements} we present the final $^{12}$C/$^{13}$C ratio at the surface. The $^{12}$C/$^{13}$C ratio can be determined observationally and is a key observational constraint for stellar models. For the low-mass models, the $^{12}$C/$^{13}$C ratio increases as TDU brings $^{12}$C synthesised from He-burning to the surface. The 3~$\Msun$ model has very inefficient HBB which results in a final $^{12}$C/$^{13}$C ratio of 89. As a result of the CN cycle during HBB the $^{12}$C/$^{13}$C ratio for the intermediate-mass models reaches an equilibrium value of approximately 3.  This is in agreement with the value found by \citet{1998A&A...332L..17F}. The final $^{12}$C/$^{13}$C ratio is greater than 3 as it increases as a result of HBB ceasing while TDU continues. 

All the models become carbon-rich with the C/O ratio increasing to above unity. The low-mass models have a high final C/O ratio. This is because the surface abundance of O only increases slightly compared to the increase of C. Despite the more massive models having a higher $T_{\rm bce}$ and more efficient HBB, the 4~$\Msun$ model has the lowest C/O ratio of 3.38 of all the intermediate-mass models.

The final surface abundance of N is reasonably constant for the low-mass models, with the increase in [N/Fe] to $\sim 0.5$ a result of FDU. The N yield increases with increasing initial mass from $1.95 \times 10^{-5}$~$\Msun$ for the 1~$\Msun$ model to $2.34 \times 10^{-4}$~$\Msun$ for the 3~$\Msun$ model. The intermediate-mass models experience HBB and, as such, have a higher final surface abundance of N compared to the low-mass models. The 4~$\Msun$ model has the highest [N/Fe] ratio (with a value of $2.62$) and N yield ($5.94 \times 10^{-2}$~$\Msun$) of all the models.  This is a result of the 4~$\Msun$ model having the largest amount of material brought to the surface, which provides additional primary C to be converted to N.

The low-mass models between 2 and 2.5~$\Msun$ have a similar final surface abundance of O, with an [O/Fe] ratio of around $0.3$. The final O abundance is lower for the intermediate-mass models due to its destruction from HBB. For the models with a mass of 4.5~$\Msun$ and higher, the [O/Fe] abundances are negative, down to $-0.64$ for the 7~$\Msun$ model. As expected, the 7~$\Msun$ model has the lowest net yield of O with $-2.20 \times 10^{-3}$~$\Msun$. The highest yield of $6.34 \times 10^{-4}$~$\Msun$ occurs for the 2.5~$\Msun$ model.

There is only one stable isotope of F ($^{19}$F) which is produced through the $^{15}$N($\alpha$,$\gamma$)$^{19}$F reaction in the He-intershell \citep{1992A&A...261..164J,1996A&A...311..803M,Lugaro:2004en,2009ApJ...694..971A}. The F synthesised during the preceding TP is mixed to the surface during TDU. The final F surface abundance increases for each model, with [F/Fe] up to 2.27 for the 2.5~$\Msun$, before decreasing to sub-solar values for the 6 and 7~$\Msun$ models. The decrease in F is caused by the destruction of $^{19}$F through $\alpha$ capture to produce $^{22}$Ne.  In the more massive models, the F yield also decreases as temperatures during HBB allow for the destruction of F to take place via the $^{19}$F($p$,$\alpha$)$^{16}$O reaction. The 5, 5.5, 6, and 7~$\Msun$ models have a negative F yield, with the lowest net yield of $-1.41 \times 10^{-7}$~$\Msun$ occurring for the 7~$\Msun$ model.

\begin{figure}
\begin{center}
\includegraphics[width=\columnwidth]{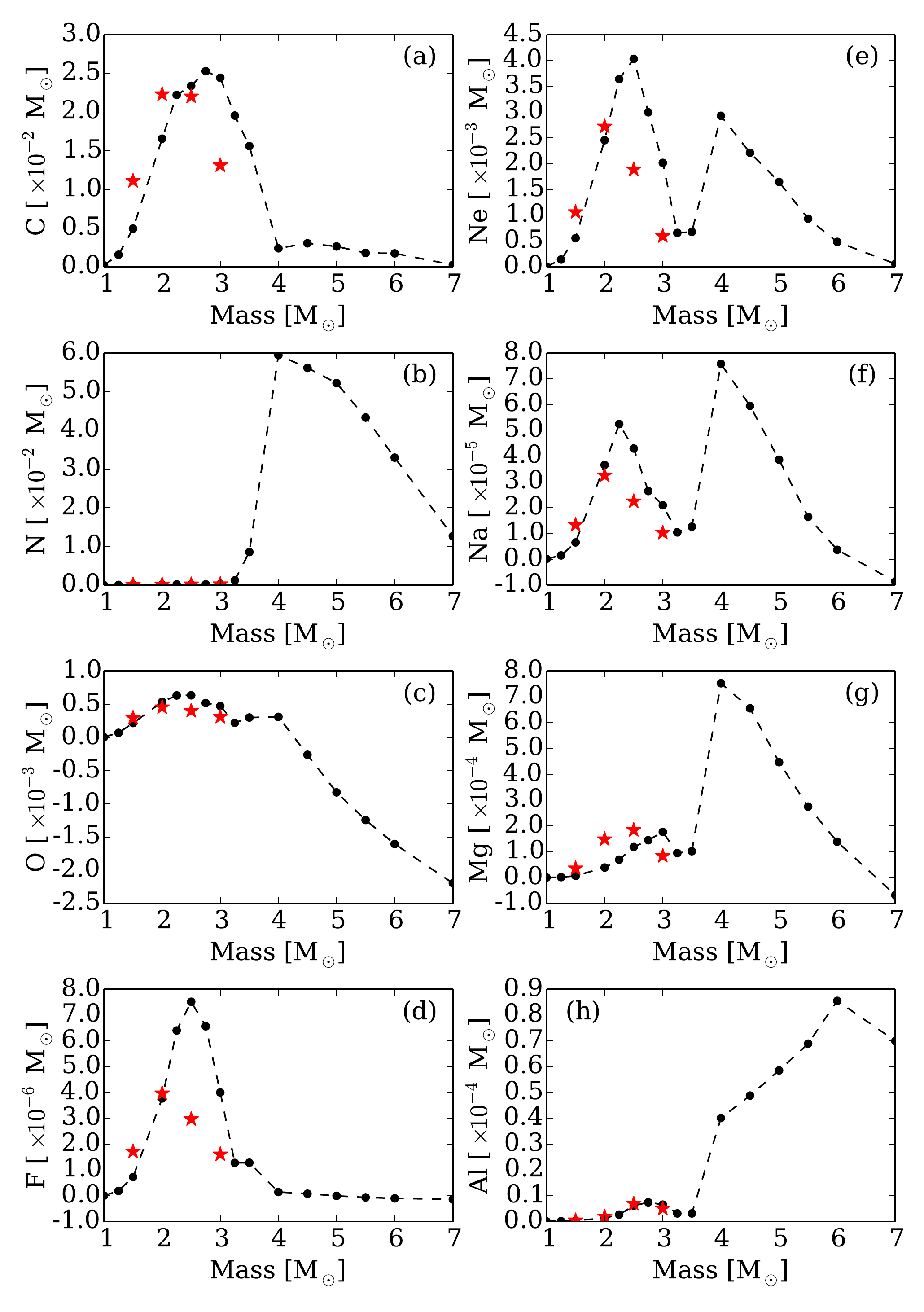} 
\caption{Net yields of select elements lighter than Si as a function of initial mass. Results from \citet{2011ApJS..197...17C} are shown as red stars.}
\label{fig:f11}
\end{center}
\end{figure}

\subsubsection{Ne, Na, Mg, and Al}

Only Ne and Na are noticeably affected by FDU and SDU whereas the surface abundances of Mg and Al  do not change during FDU for the low-mass models and changes by only 35 per cent for $^{27}$Al for the intermediate-mass models. Of all the models, the 2.25~$\Msun$ has the highest final surface abundance of [Ne/Fe] and [Na/Fe]  as a result of FDU and TDU with 1.41 and 1.29, respectively (Table~\ref{tab:lightelements}). The Ne-Na chain is activated during HBB for the intermediate-mass models and the main result of the Ne-Na chain is the production of $^{23}$Na, the only stable isotope of Na, at the expense of $^{22}$Ne. The 4~$\Msun$ model has the highest yield of Na with $7.57 \times 10^{-5}$~$\Msun$. The 7~$\Msun$ model is the only model with a negative Na yield.

The highest final surface abundances of Mg and Al occur for the 4 and 6~$\Msun$ models, respectively. This is also reflected in the net yields. The yield of Mg increases with increasing initial mass before decreasing once the peak yield of $7.53 \times 10^{-4}$~$\Msun$ is reached for the 4~$\Msun$ model. The 3.25 and 3.5~$\Msun$ models are exceptions to this trend. This is because the models (compared to lower and higher mass models) have less TDU, the absence of a PMZ resulting in fewer neutron captures onto $^{24}$Mg, lower activation of the $^{22}$Ne + $\alpha$ reaction compared to intermediate-mass models, and the Mg-Al chain is not activated. The production of Al increases with increasing mass for the intermediate-mass models however the 7~$\Msun$ model has a lower Al yield than the 6~$\Msun$ due to HBB temperatures being high enough for the production of $^{28}$Si to occur at the expense of $^{27}$Al.

\subsubsection{Other light elements}

There is a slight enhancement of up to approximately $0.15$~dex for Si, S, Cl, and Ar. The 7~$\Msun$ model produces the most Si with a net yield of $7.73 \times 10^{-5}$~$\Msun$ (Table~\ref{tab:yields_ims}). [P/Fe] increases by around 0.5~dex for the intermediate-mass models with the maximum enhancement occurring for the 4~$\Msun$ model. Sc is produced with the 4~$\Msun$ model showing the largest enhancement of  [Sc/Fe] ($\approx$ $0.4$~dex). Of the Fe-group elements, Cu is enhanced the most in the 4~$\Msun$ model increasing by $1.03$~dex for [Cu/Fe]. The abundance of [Zn/H] has been proposed to be a good proxy of [Fe/H] in planetary nebulae \citep[e.g.][]{2001ApJ...562..515D,Smith:2014tm}. The low-mass models have [Zn/Fe] enhancements between 0 and 0.32~dex with the 1~$\Msun$ model having no increase and the 3~$\Msun$ model having the largest increase. The smallest enhancement for the intermediate-mass models occurs for the 7~$\Msun$ model with a [Zn/Fe] ratio of $0.22$ whereas the 4~$\Msun$ model has [Zn/Fe] increase by 0.59~dex.

\subsubsection{Comparison with \citet{2011ApJS..197...17C}}

Figure~\ref{fig:f11} also presents the net yields of \citet{2011ApJS..197...17C} for each of the initial stellar masses in common with the models presented here: 1.5, 2, 2.5, and 3~$\Msun$. For these low masses, the net yield of C (and O) follows the total amount of material mixed to the surface by TDU. The 2.5 and 3~$\Msun$ models of \citet{2011ApJS..197...17C} have a lower value of $M_{\rm TDU}$, which results in a lower C  (and O) yield in Figure~\ref{fig:f11}a; with the largest yield difference $\Delta M_i$ being $1.13 \times 10^{-2}$~$\Msun$ for the 3~$\Msun$ model. 

The net yield of N agrees with the predictions of \citet{2011ApJS..197...17C}, with the N yield increasing with increasing initial mass as a result of FDU. 
For the \citet{2011ApJS..197...17C} yields of F, Ne, and Na, the 2~$\Msun$ model has the largest net yield as a result of having the largest value of $M_{\rm TDU}$. For the models presented here, the yields peak at 2.5~$\Msun$ except for Na which peaks at 2.25~$\Msun$. The yields of Mg and Al peak at 3~$\Msun$ and 2.75~$\Msun$, respectively.

\subsection{The neutron-capture elements}

\begin{figure*}
\begin{center}
\includegraphics[width=2\columnwidth]{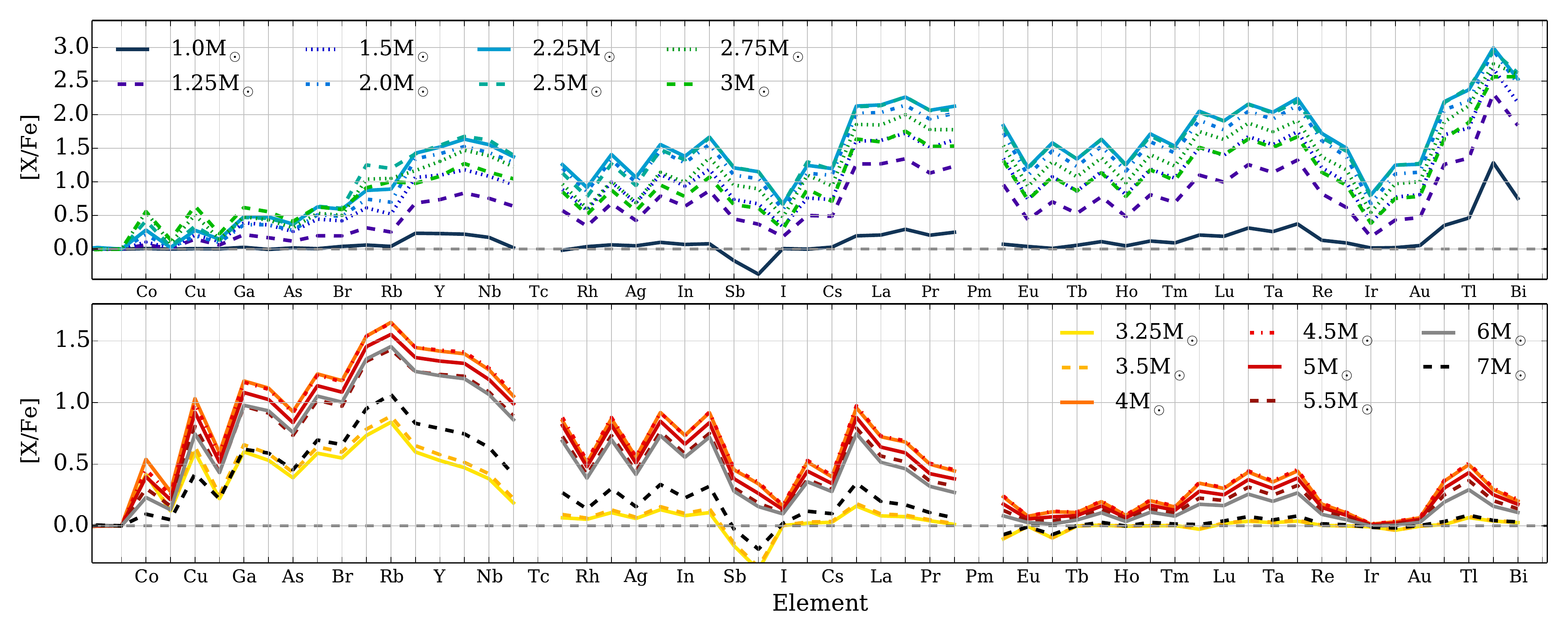} 
\caption{Final surface abundances relative to Fe for each of the models for the elements heavier than Mn. Low-mass models are presented in the top plot while intermediate-mass models are presented in the bottom plot. The elements are ordered by increasing atomic number.  For an explanation of why some abundances have a value less than solar, see the caption of Figure~\ref{fig:f4} and Section~\ref{sec:code}.}
\label{fig:f12}
\end{center}
\end{figure*}

In this section we present final surface abundances and net stellar yield predictions for selected neutron-capture elements (Rb, Sr, Y, Zr, Ba, La, Ce, and Pb). Table~\ref{tab:heavyelements} presents the final surface abundances for selected neutron-capture elements and $s$-process indicators.
The distribution of the final surface abundances [X/Fe] for these elements is shown in Figure~\ref{fig:f12}.

  \begin{table*}
   \def\arraystretch{1.2}
 \begin{center}
  \caption{Final surface abundances for select neutron-capture elements and $s$-process indicators.}
 \label{tab:heavyelements}
   \begin{tabular}{ccccccccccc}
  \tableline\tableline
Mass & [Rb/Fe] & [Sr/Fe] & [Y/Fe] & [Zr/Fe] & [Ba/Fe] & [La/Fe] & [Ce/Fe] & [Pb/Fe] & [$ls$/Fe] & [$hs$/Fe] \\
\hline
 1.00 & 0.04 & 0.24 & 0.23 & 0.22 & 0.20 & 0.21 & 0.29 & 1.28 & 0.23 & 0.18   \\
1.25 & 0.25 & 0.68 & 0.74 & 0.83 & 1.27 & 1.27 & 1.34 & 2.31 & 0.75 & 1.18   \\
1.50 & 0.52 & 1.07 & 1.10 & 1.18 & 1.60 & 1.62 & 1.72 & 2.68 & 1.11 & 1.55   \\
2.00 & 0.70 & 1.34 & 1.42 & 1.53 & 2.02 & 2.03 & 2.14 & 2.95 & 1.43 & 1.95   \\
2.25 & 0.89 & 1.43 & 1.52 & 1.64 & 2.13 & 2.14 & 2.26 & 2.99 & 1.53 & 2.06   \\
2.50 & 1.20 & 1.42 & 1.54 & 1.68 & 2.12 & 2.13 & 2.26 & 2.94 & 1.55 & 2.03   \\
2.75 & 1.05 & 1.17 & 1.30 & 1.47 & 1.85 & 1.85 & 2.00 & 2.75 & 1.31 & 1.75   \\
3.00 & 1.00 & 0.98 & 1.08 & 1.27 & 1.64 & 1.59 & 1.75 & 2.56 & 1.11 & 1.51   \\
3.25 & 0.84 & 0.60 & 0.53 & 0.47 & 0.16 & 0.08 & 0.07 & 0.04 & 0.53 & 0.03   \\
3.50 & 0.89 & 0.65 & 0.58 & 0.51 & 0.18 & 0.10 & 0.09 & 0.04 & 0.58 & 0.04   \\
4.00 & 1.65 & 1.45 & 1.42 & 1.40 & 0.96 & 0.72 & 0.68 & 0.29 & 1.41 & 0.58   \\
4.50 & 1.65 & 1.45 & 1.43 & 1.41 & 0.97 & 0.73 & 0.69 & 0.31 & 1.42 & 0.59   \\
5.00 & 1.55 & 1.37 & 1.34 & 1.32 & 0.88 & 0.64 & 0.59 & 0.25 & 1.33 & 0.51   \\
5.50 & 1.43 & 1.25 & 1.23 & 1.21 & 0.79 & 0.57 & 0.52 & 0.20 & 1.23 & 0.44   \\
6.00 & 1.45 & 1.25 & 1.22 & 1.19 & 0.75 & 0.52 & 0.46 & 0.16 & 1.22 & 0.40   \\
7.00 & 1.07 & 0.83 & 0.79 & 0.75 & 0.35 & 0.20 & 0.17 & 0.04 & 0.79 & 0.13   \\
\tableline
  \end{tabular} 
\\ 
 \end{center}
\end{table*}

As illustrated in Figure~\ref{fig:f12}, the low-mass models produce a final surface abundance distribution of the neutron-capture elements that has peaks at Sr, Zr, Ba, and Pb as discussed for the case of the 2~$\Msun$ model. In comparison the intermediate-mass models produce a peak at Rb. This difference is due to a combination of the addition of a PMZ for the low-mass models and the activation of the $^{22}$Ne neutron source for the intermediate-mass models.

The final surface abundance of Rb increases with increasing initial mass for the low-mass models, up to $1.20$ for the 2~$\Msun$ model before decreasing slightly for the 2.75 and 3~$\Msun$ models. This increase is mainly the result of the mild activation of the $^{22}$Ne neutron source.  For the intermediate-mass models, the highest final surface abundance and yield of all the neutron-capture elements occurs for Rb where branching points are activated and the total neutron exposure is lower than in the low-mass models. The highest final surface abundance for Rb occurs for the 4 and 4.5~$\Msun$ models where both have a final [Rb/Fe] value of $1.65$. The 4.5~$\Msun$ model has the highest yield of Rb with $1.64 \times 10^{-7}$~$\Msun$.

Each model has a similar final abundance for [Sr/Fe], [Y/Fe], and [Zr/Fe], three first $s$-process peak elements. Of the three elements,  Sr has the lowest final abundance and Zr has the highest for each of the low-mass models. For the intermediate-mass models, the trend is reversed with Sr having the highest abundance and Zr, the lowest of the three elements. Of the intermediate-mass models, the 4.5~$\Msun$ model has the highest abundances with 1.45, 1.43, and 1.41 for Sr, Y, and Zr, respectively. This is also reflected in the yields of the 4.5~$\Msun$ model where it has the highest Sr and Y yields of all the models, $2.38 \times 10^{-7}$ and $4.85 \times 10^{-8}$~$\Msun$, respectively. For Zr, the 2.5~$\Msun$ model has the highest yield of $1.19 \times 10^{-7}$~$\Msun$.

For the low-mass models, the final surface abundances of Ba, La, and Ce are higher than the abundances of Sr, Y, and Zr. The 2.25~$\Msun$ model has the highest final [Ba/Fe], [La/Fe], and [Ce/Fe] values of all the models (see Table~\ref{tab:heavyelements}). The abundances of Ba, La, and Ce for the intermediate-mass models never reach above enhancements of 1 dex.  The 4.5~$\Msun$ model once again has the highest final abundances of the intermediate-mass models for these three elements whereas the 7~$\Msun$ has the lowest. Of all the models, the 2.5~$\Msun$ model has the highest yields for Ba, La, and Ce with $2.23 \times 10^{-7}$, $2.28 \times 10^{-8}$, and $8.00 \times 10^{-8}$~$\Msun$, respectively.

\begin{figure}
\begin{center}
\includegraphics[width=\columnwidth]{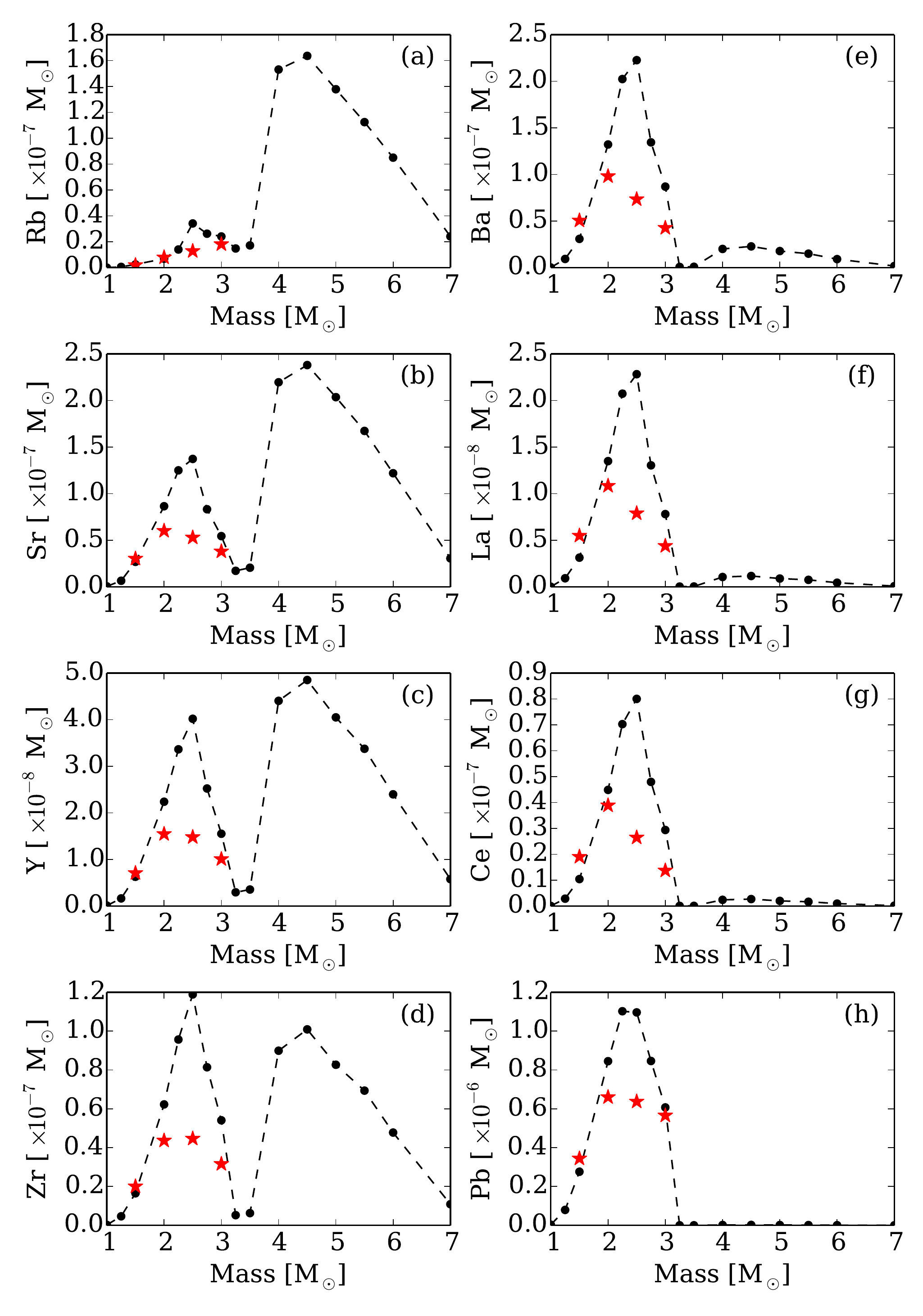} 
\caption{Net yields of select neutron-capture elements as a function of initial mass. Results from \citet{2011ApJS..197...17C} are shown as red stars.}
\label{fig:f13}
\end{center}
\end{figure}

The low-mass models produce more Pb compared to the intermediate-mass models (Figure~\ref{fig:f12}). Once the first and second $s$-process peaks reach equilibrium, any increase in their abundance is prevented and only the abundance of Pb increases. The value of [Pb/Fe] reaches a maximum value of $2.99$ for the 2.25~$\Msun$ model. The intermediate-mass models produce minimal Pb as discussed in the case of the 5~$\Msun$ model, with the final surface abundance ranging from 0.04 for the 3.25~$\Msun$ model to 0.31 for the 4.5~$\Msun$ model. For the low-mass models, the Pb yield increases with increasing mass before reaching a plateau of approximately $1.7 \times 10^{-6}$~$\Msun$ for the 2.25 to 2.5~$\Msun$ models (see Figure~\ref{fig:f13}). The Pb yield then drops below approximately a few times $10^{-9}$~$\Msun$ for the intermediate-mass models.

\begin{figure}
\begin{center}
\includegraphics[width=\columnwidth]{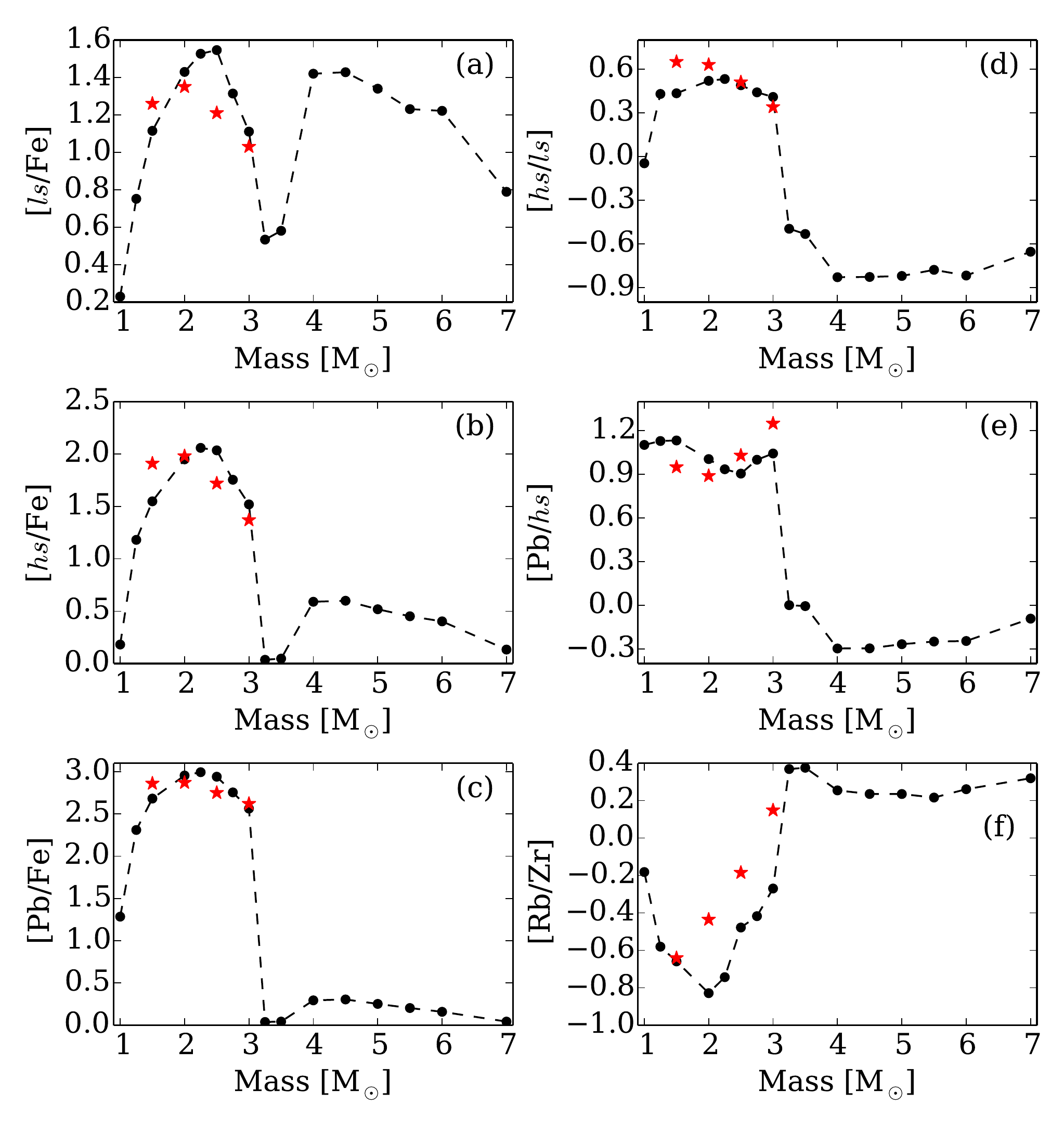} 
\caption{Distribution of [$hs$/$ls$] and [Pb/$hs$] with initial mass. Results from \citet{2011ApJS..197...17C} are shown as red stars.}
\label{fig:f14}
\end{center}
\end{figure}

Figure~\ref{fig:f14} presents the distribution of the final surface abundance of the $s$-process indicators [$ls$/Fe], [$hs$/Fe], [Pb/Fe], [$hs$/$ls$], [Pb/$hs$], and [Rb/Zr] with initial mass. The [$ls$/Fe] ratio increases with increasing initial mass for the low-mass models up to 2.5~$\Msun$ before decreasing for the 2.75, 3, and 3.25~$\Msun$ models. The [$ls$/Fe] ratio then increases up to $1.43$ for the 4.5~$\Msun$ model before decreasing again with increasing initial mass. The [$hs$/Fe] and [Pb/Fe] values for the intermediate-mass models are less than $0.6$.

Figures~\ref{fig:f14}d and ~\ref{fig:f14}e illustrate the trend of the $s$-process indicators [$hs$/$ls$] and [Pb/$hs$] with initial mass. These intrinsic $s$-process indicators are independent of the amount of TDU and help constrain the neutron source and neutron exposure for the $s$-process. The low-mass models, excluding the 1~$\Msun$ model, plateau at approximately $0.5$ for [$hs$/$ls$] while the intermediate-mass plateau at a sub-solar value of around $-0.8$. For [Pb/$hs$], the low-mass models fluctuate between $\sim 0.9$ and $1.2$. The intermediate-mass models have a sub-solar value of approximately $-0.2$ with the 3.25 and 3.5~$\Msun$ models having a value close to solar. 

Figure~\ref{fig:f14}f illustrates the trend of the final surface [Rb/Zr] ratio with initial mass, where Rb and Zr are both first peak neutron-capture elements. This ratio is an indicator of the neutron density with a positive ratio resulting from higher densities produced by the $^{22}$Ne neutron source. The intermediate-mass models show a fairly constant [Rb/Zr] ratio; between 0.2 and 0.4~dex. The low-mass models, however, first decrease with increasing initial mass from $-0.2$ for the 1~$\Msun$ model to $-0.8$ for the 2~$\Msun$ model, then increase to approximately $-0.3$~dex for the 3~$\Msun$ model. The increase in [Rb/Zr] is due to temperatures increasing in the pulse-driven convective zone so that the $^{22}$Ne neutron source is mildly activated.

\subsubsection{Comparison with \citet{2011ApJS..197...17C}}

 Figure~\ref{fig:f13} also presents the yield predictions from \citet{2011ApJS..197...17C} for the 1.5, 2, 2.5, and 3~$\Msun$ models. With the exception of Rb and Zr, the \citet{2011ApJS..197...17C} yields for the $s$-process elements shown in Figure~\ref{fig:f13} have the highest value for the 2~$\Msun$, which has the largest $M_{\rm TDU}$ value of the four models.  In contrast, the yield predictions of the models presented here peak at 2.5~$\Msun$ except for Pb which peaks at 2.25~$\Msun$. The yield of Pb reaches a plateau between 2.25~$\Msun$ and 2.5~$\Msun$. The largest difference in the yield predictions occurs for the 2.5~$\Msun$ model.

Figure~\ref{fig:f14} presents $s$-process indicator predictions for the \citet{2011ApJS..197...17C} models. The $s$-process indicators show a similar trend with mass between the  \citet{2011ApJS..197...17C} models and the models presented here. The [$ls$/Fe], [$hs$/Fe], and [Pb/Fe] increase with increasing mass before reaching a peak and decreasing.  The low-mass models presented here show a flat distribution for [$hs$/$ls$] of approximately 0.5~dex whereas the [$hs$/$ls$] predictions of \citet{2011ApJS..197...17C} decrease with increasing initial mass, from 0.65~dex to 0.34~dex. The values of [Pb/$hs$] fluctuate between $\sim 0.9$ and $1.2$ however the models of \citet{2011ApJS..197...17C} have [Pb/$hs$] mostly increasing with increasing initial mass. The [Rb/Zr] ratios are higher for the \citet{2011ApJS..197...17C} models, due to higher peak temperatures during TPs (see Table~\ref{tab:prop}). Note that the rate of the $^{22}$Ne source we use from \citet{2010NuPhA.841...31I} is comparable to the \citet{2001PhRvL..87t2501J} rate used by \citet{2011ApJS..197...17C}. The final abundances of [Zr/Fe] are lower for the \citet{2011ApJS..197...17C} models (excluding the 1.5~$\Msun$ model) further increasing the final [Rb/Zr] ratio.

\section{Effects of varying the mass of the PMZ}
\label{sec:pmz}

The extent in mass of the PMZ and the profile of the proton abundance in the PMZ are unknown parameters which introduce additional uncertainty into the elemental abundances and stellar yields for the low-mass AGB stellar models \citep{2009PASA...26..133S}. Here, we investigate the effect of varying the extent in mass of the PMZ, while keeping fixed the exponential profile of the proton abundance. We have computed the 3~$\Msun$ model using three different values for the extent in mass of the PMZ: (0.5, 1, 2) $\times~10^{-3}$~$\Msun$ as well as a model without the inclusion of a PMZ. 

\begin{figure}
\begin{center}
\includegraphics[width=\columnwidth]{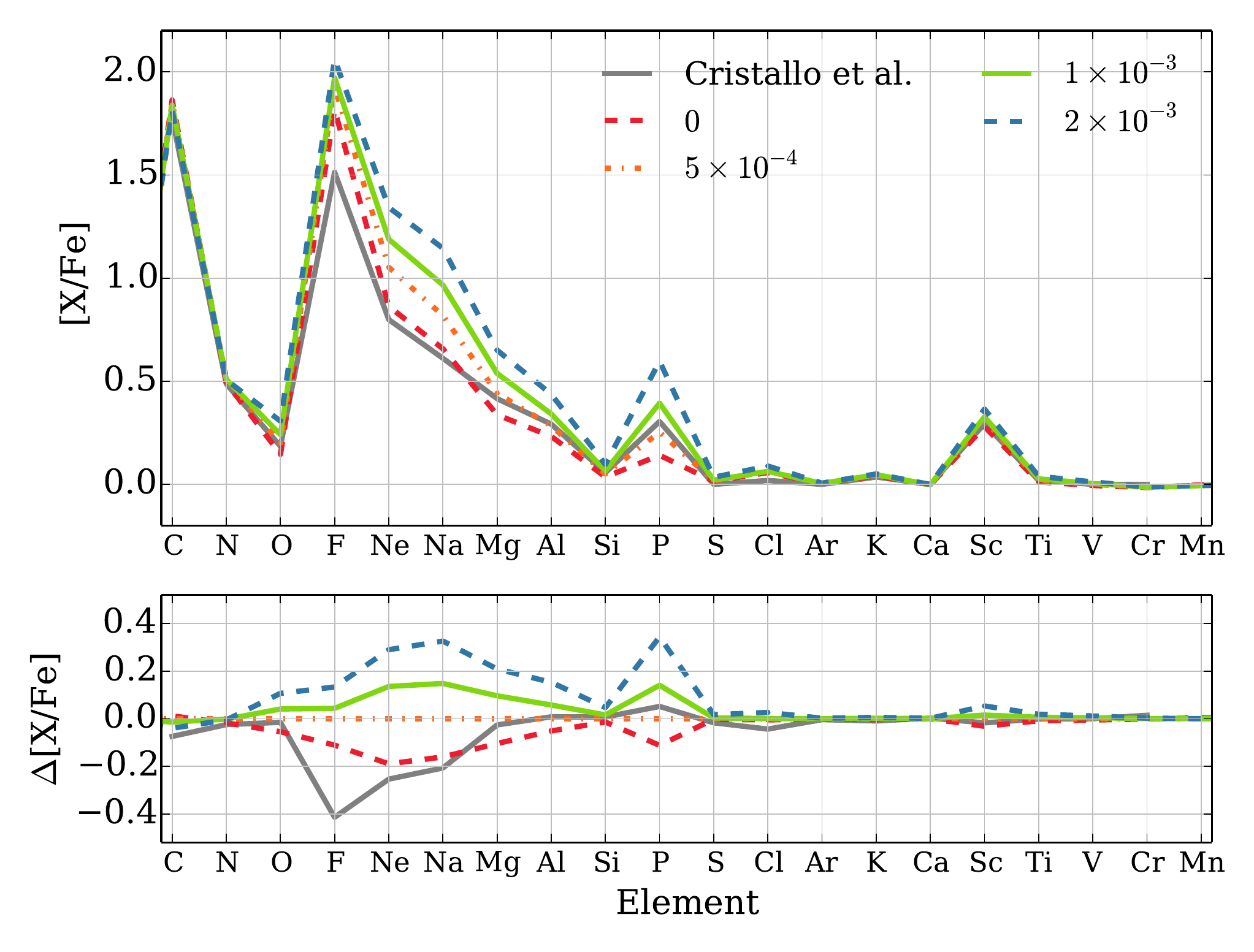} 
\caption{Final surface abundance ratios for each 3~$\Msun$ model with a different PMZ mass (in units of $\Msun$ as indicated in the legend) for the elements lighter than Fe. The final surface abundances for the 3~$\Msun$ model of \citet{2011ApJS..197...17C} are also presented. The bottom panel illustrates the absolute difference between each model and the standard PMZ mass, $5 \times 10^{-4}$~$\Msun$. The elements are ordered by increasing atomic number.}
\label{fig:f15}
\end{center}
\end{figure}

The difference in the final surface abundances of the light elements compared to the standard PMZ mass of $5 \times 10^{-4}$~$\Msun$ is shown in Figure~\ref{fig:f15}. The model with a PMZ of $2 \times 10^{-3}$~$\Msun$ has the largest increase in [X/Fe] for elements lighter than Fe (excluding C) with respect to the standard case; the largest increase is exhibited by Ne, Na, and P where $\Delta$[Ne/Fe], $\Delta$[Na/Fe], and $\Delta$[P/Fe] are approximately $+0.3$. Between the model with the standard PMZ and the \citet{2011ApJS..197...17C} model, the [F/Fe] ratio shows the largest difference of $0.4$~dex. The final abundances of [Ne/Fe] and [Na/Fe] are also lower by approximately $0.2$~dex in the Cristallo et al. model compared to our standard PMZ case. 

As a larger PMZ extends over a larger mass range in the intershell, it reaches into regions of higher temperature. The higher temperatures cause the $^{13}$C pocket to form sooner and deeper in the intershell for the models with a more massive PMZ compared with the standard PMZ mass. The larger PMZ also results in larger $^{13}$C and  $^{14}$N pockets forming in the intershell. The extra $^{14}$N is captured by $\alpha$ particles during subsequent TPs to produce $^{22}$Ne. The increases in Ne and Na are therefore the result of the increased production of $^{22}$Ne, where the $^{22}$Ne is dredged to the surface. Some of the newly synthesised $^{22}$Ne is captured by protons in the H-shell during the next interpulse period to make extra $^{23}$Na. When compared to the standard case, the model without a PMZ produces lower abundances with [Ne/Fe] and [Na/Fe] showing deficiencies of approximately $-0.2$~dex.

Of the elements between Si and Mn, only P shows a non-negligible production due to the increase in the mass of the PMZ. There is only one stable isotope of P ($^{31}$P) and it can be produced through neutron capture in AGB stars. The increase in [P/Fe] with increasing PMZ mass is due to the increased number of neutrons available for neutron capture. 

\begin{figure}
\begin{center}
\includegraphics[width=\columnwidth]{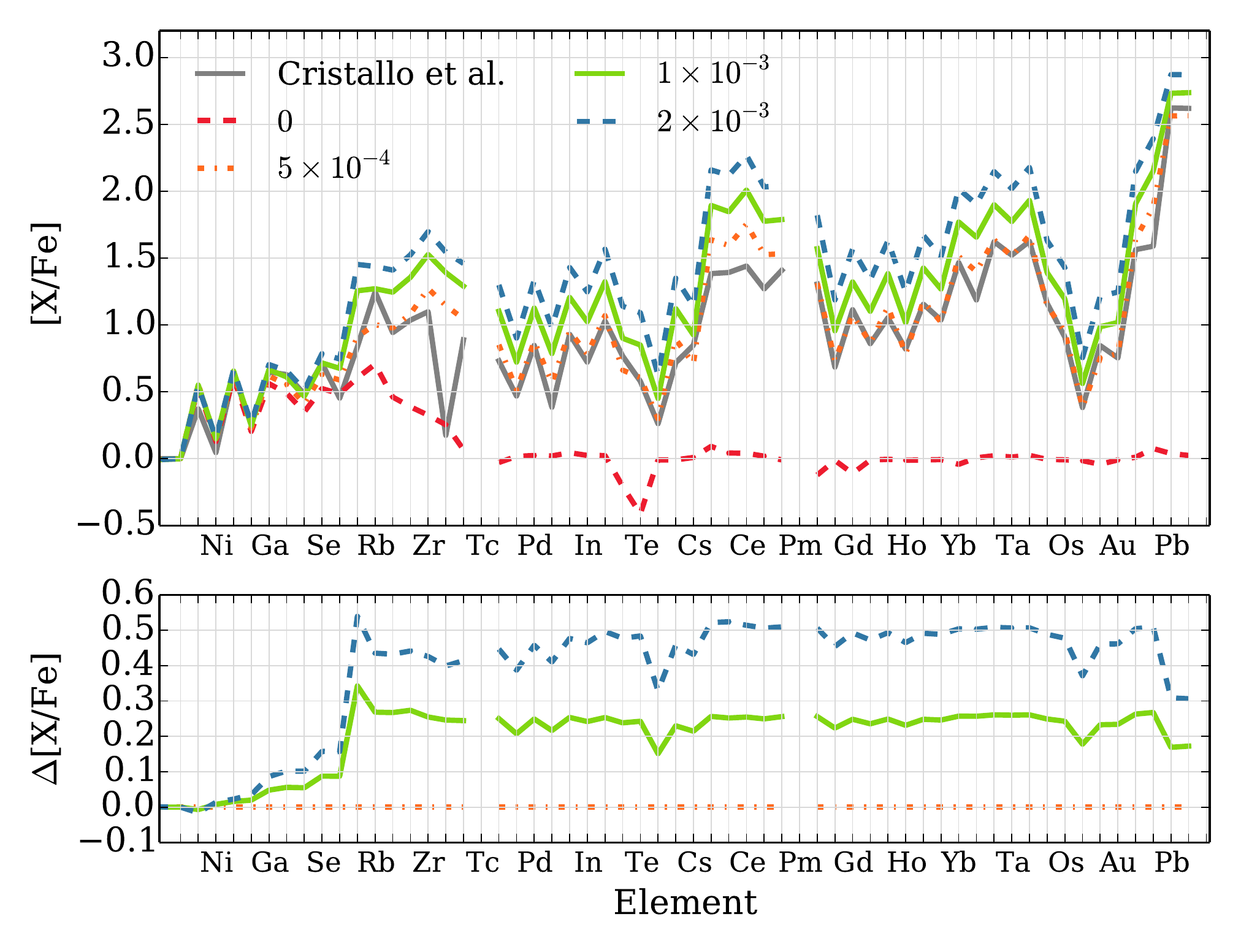} 
\caption{Final surface abundance ratios for each 3~$\Msun$ model with a different PMZ mass (in units of $\Msun$ as indicated in the legend) for the elements heavier than Fe. The final surface abundances for the 3~$\Msun$ model of \citet{2011ApJS..197...17C} are also presented. The elements are ordered by increasing atomic number. For an explanation of why some abundances have a value less than solar, see the caption of Figure~\ref{fig:f4} and Section~\ref{sec:code}.}
\label{fig:f16}
\end{center}
\end{figure}

The effect of the PMZ mass on the final surface abundance distribution for the neutron-capture elements is illustrated in Figure~\ref{fig:f16}.  The height of the abundance peaks increases with increasing PMZ mass with the general shape of the distribution of the $ls$ and $hs$ elements remaining the same.  
The higher temperatures reached by a more massive PMZ increases the rate of the $^{13}$C($\alpha$,$n$)$^{16}$O reaction resulting in a higher peak neutron density. This has the effect of increasing the efficiency of the branching points at $^{85}$Kr and $^{86}$Rb producing more $^{86}$Kr and $^{87}$Rb, both of which have a magic number of neutrons. Figure~\ref{fig:f16} shows that [Kr/Fe] has the largest increase between the model with a PMZ mass of $2 \times 10^{-3}$~$\Msun$ and the model with the standard PMZ.

We attribute the smaller increase in the Pb abundance with increasing PMZ mass (compared to the $ls$ and $hs$ elements) to the lower neutron exposure experienced in each $^{13}$C pocket  \citep[see e.g.,][]{1998ApJ...497..388G}. Compared to the model with the standard PMZ mass, the neutron exposures for models with a more massive PMZ become increasingly lower with each interpulse period.
Another result of increased $^{13}$C burning temperatures is that the $^{13}$C nuclei are consumed faster and the duration of the $^{13}$C pocket is shorter for the models with a more massive PMZ. 

When a PMZ is not added in the post-processing nucleosynthesis calculations, the effect of the $^{22}$Ne neutron source is more evident. In this case, the largest final abundance occurs for [Rb/Fe] due to branching points opening in the $s$-process path at Rb. The much lower neutron exposure however implies minimal production of second $s$-process peak elements and Pb.

In contrast to our models with a PMZ, the model of \citet{2011ApJS..197...17C} has Rb as the most enhanced first $s$-process peak element. However, our 3~$\Msun$ model (with the standard PMZ mass) has a slightly higher Rb yield than the \citet{2011ApJS..197...17C} model, with a net yield of $2.41 \times 10^{-8}$~$\Msun$ compared to $1.81 \times 10^{-8}$~$\Msun$. This is due to the faster increase in Rb with TP number where [Rb/Fe] asymptotically approaches $0.8$ and more of this enriched material is then ejected through mass loss.
The \citet{2011ApJS..197...17C} model has a lower abundance of second $s$-process peak elements but a higher Pb abundance than the model with the standard PMZ mass of $5 \times 10^{-4}$~$\Msun$. 

\begin{figure}
\begin{center}
\includegraphics[width=\columnwidth]{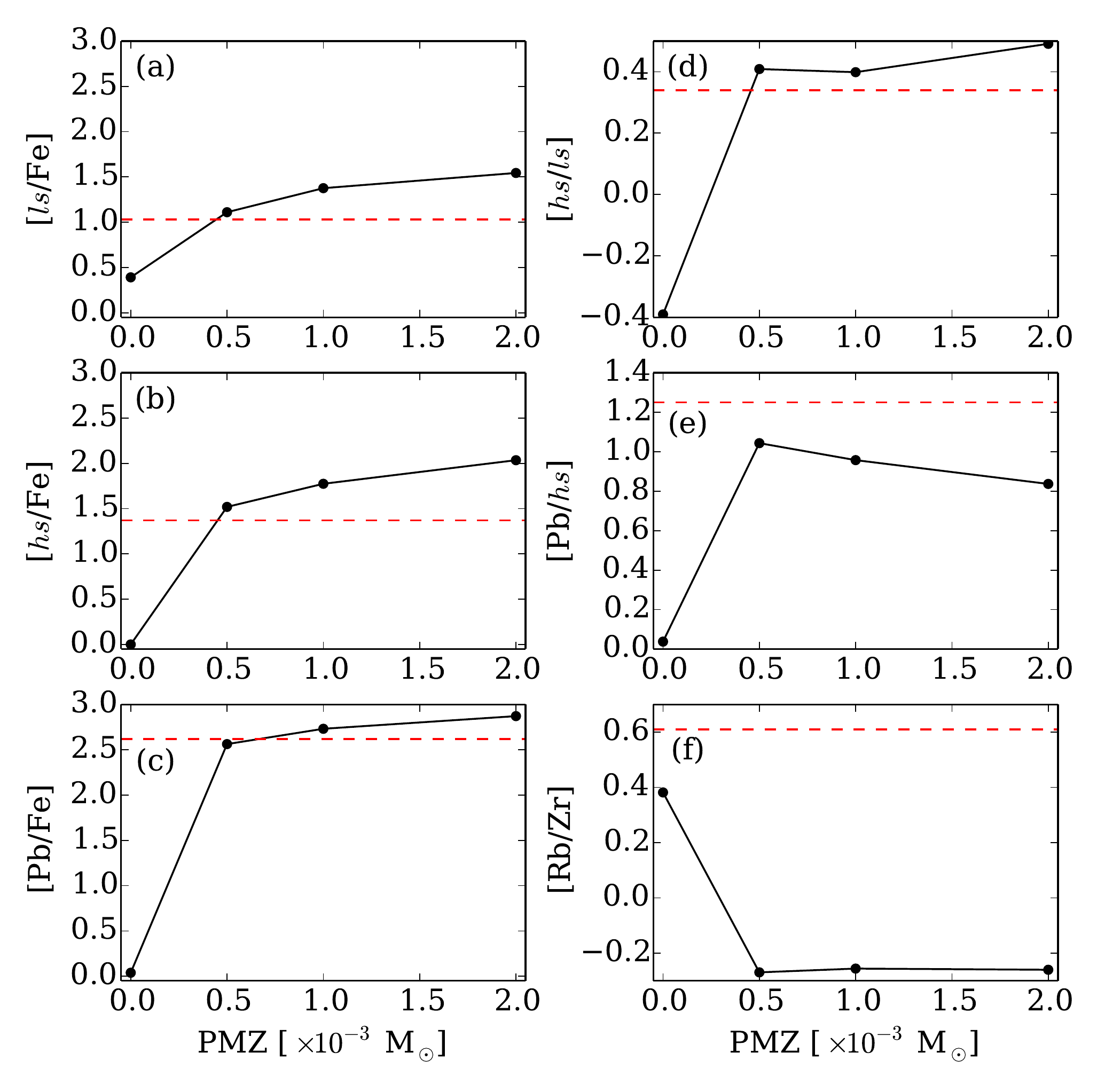} 
\caption{Distribution of [$ls$/Fe], [$hs$/Fe], [Pb/Fe], [$hs$/$ls$], [Pb/$hs$], and [Rb/Zr] with varying PMZ mass for each 3~$\Msun$ model showing the behaviour of the $s$-process peaks. The results from \citet{2011ApJS..197...17C} are shown as a horizontal dashed line.}
\label{fig:f17}
\end{center}
\end{figure}

All the models with a PMZ produce relatively high abundances of neutron-capture elements (Figure~\ref{fig:f17}) and the values of [$ls$/Fe], [$hs$/Fe], and [Pb/Fe] increase with increasing PMZ mass. When comparing the model with a PMZ of $2 \times 10^{-3}$~$\Msun$ to the model with the standard PMZ, [$ls$/Fe] increases by $0.43$~dex while for [$hs$/Fe] the increase is $0.52$~dex. For [Pb/Fe] there is an increase of $0.3$~dex. 

Figure~\ref{fig:f17} also highlights the effect of changing the mass of the PMZ on the intrinsic $s$-process indicators [$hs$/$ls$] and [Pb/$hs$]. For the models with a PMZ, there is an absolute difference of only 0.08 for the [$hs$/$ls$]. The small change in [$hs$/$ls$] is due to the abundances reaching equilibrium \citep[see][]{2012ApJ...747....2L}. For [Pb/$hs$] there is a decrease of 0.21 dex when increasing the PMZ mass from $5 \times 10^{-4}$ to $2 \times 10^{-3}$~$\Msun$. This is a result of the lower neutron exposure when the PMZ mass is higher.

Despite the different approaches, there is a reasonable agreement between the two groups, as testified by the $s$-process indicators reported in Figure~\ref{fig:f17}. However, there is a disagreement between the final abundance of [Rb/Zr] between the \citet{2011ApJS..197...17C} model and our models with a PMZ. The models presented here have a sub-solar [Rb/Zr] ratio of approximately $-0.25$ whereas the \citet{2011ApJS..197...17C} model has a ratio of $\sim 0.6$ due to the higher predicted Rb abundance. The model without a PMZ is the only model that shows a [Rb/Zr] ratio above solar which is a consequence of the $^{22}$Ne($\alpha$, $n$)$^{25}$Mg reaction being the only source of neutrons.

Using our method for including a $^{13}$C pocket, it is difficult to select an appropriate mass (and profile) for the PMZ in models in the transition phase between low- and intermediate-mass. The models of \citet{2009ApJ...696..797C,2011ApJS..197...17C} use convective boundary mixing with an exponential decline in velocity  to handle the discontinuity in the radiative gradient due to the abrupt change in opacities due to the TDU episodes. This leads to a deeper TDU and to protons being partially mixed into the core. The formation of a $^{13}$C pocket then follows. Such a treatment of convective  boundary mixing results in deeper TDU relative to our models. The mixing of protons inwards in mass makes use of a free parameter $\beta$, with higher values of $\beta$ resulting in more efficient TDU. However, the effective mass of the $^{13}$C pocket does not increase with increasing values of $\beta$. The mass of the $^{13}$C pocket is at its largest when $\beta = 0.1$.  A lower or higher value of $\beta$ results in a lower abundance of neutron-capture elements. 
Our PMZ, which is added during post-processing calculations, assumes a constant mass for the proton profile at each TDU episode. In contrast, the  $^{13}$C pockets of  \citet{2009ApJ...696..797C,2011ApJS..197...17C} reduce in mass along the AGB, following the progressive shrinking in mass of the He-intershell. 

\section{Comparison to post-AGB stars}
\label{sec:pagb}

\begin{figure}
\begin{center}
\includegraphics[width=\columnwidth]{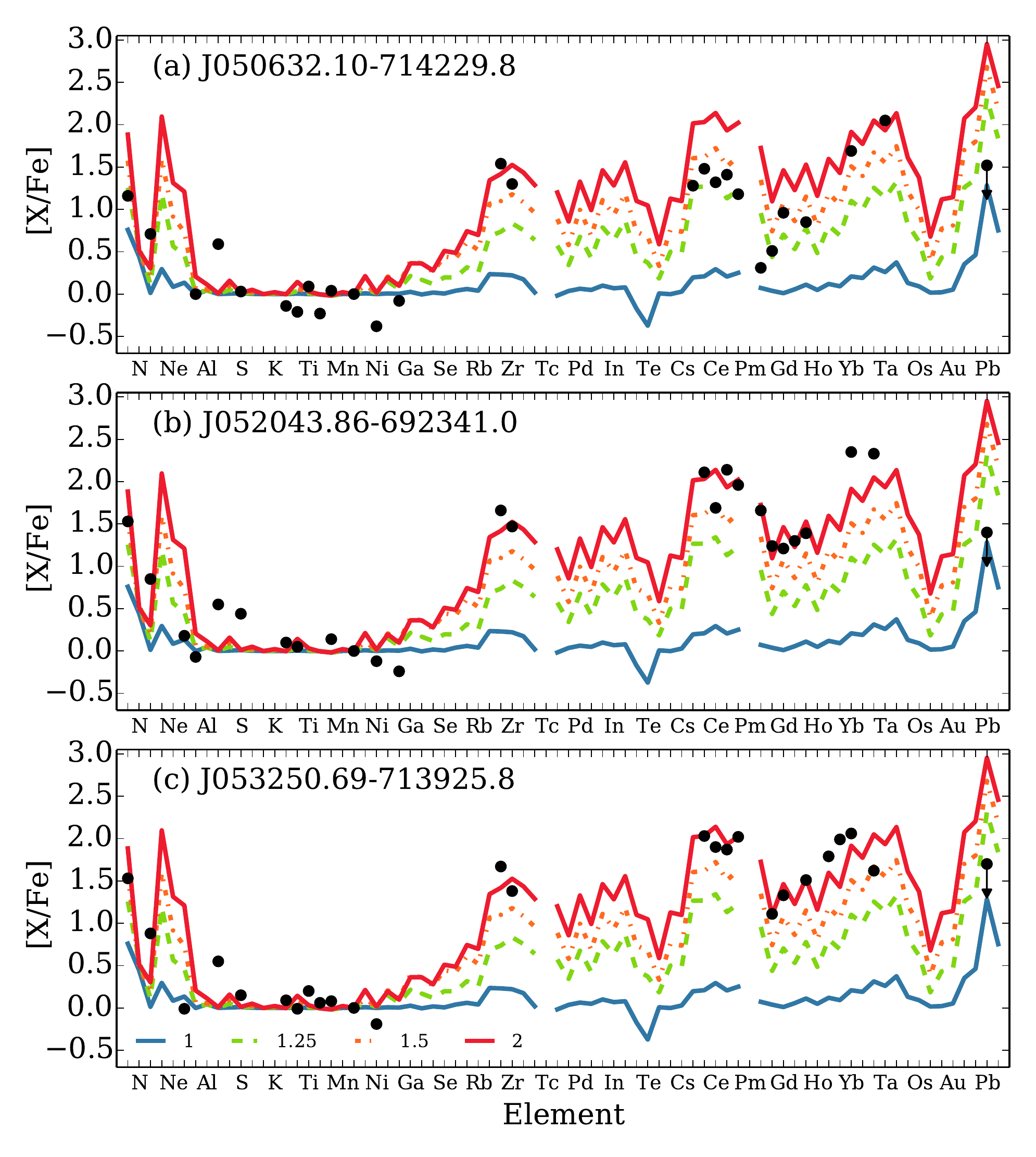} 
\caption{Comparison of three post-AGB stars \citep{2013A&A...554A.106V,2014A&A...563L...5D} to four low-mass AGB models (1, 1.25, 1.5, and 2~$\Msun$). The legend (in units of $\Msun$) is shown in the bottom panel. The post-AGB abundances are shown as black points with the Pb abundance being only an upper limit.}
\label{fig:f18}
\end{center}
\end{figure}

We compare the final surface abundance predictions to three $s$-process rich post-AGB stars in the Large Magellanic Cloud \citep{2013A&A...554A.106V,2014A&A...563L...5D}: J050632.10-714229.8, J052043.86-692341.0, and J053250.69-713925.8. The post-AGB stars have a metallicity of [Fe/H] $\approx -1.2$ and their initial masses are between 1 and 1.5~$\Msun$ \citep{2013A&A...554A.106V}. In Figure~\ref{fig:f18} we present the abundances determined by \citet{2013A&A...554A.106V} with upper limits of the Pb abundance from \citet{2014A&A...563L...5D} and the predicted final surface abundances from the models between 1 and 2~$\Msun$. 

For J052043 and J053250, the 2~$\Msun$ model is	 the best match to the neutron-capture abundances of the $ls$ and $hs$ elements. These initial masses are higher than the 1 to 1.5~$\Msun$ estimated by \citet{2013A&A...554A.106V}. The abundances of the $ls$ elements for J050632 also match the 2~$\Msun$ predictions however the $hs$ elements are better matched by the 1.25 or 1.5~$\Msun$ models.

As noted by \citet{2014A&A...563L...5D}, the observed upper limits of the Pb abundance are well below the predicted values. This is in conflict with calculations of AGB models including those presented here. Model predictions of low-metallicity AGB stars suggest that the Pb abundance should be higher than that of the second $s$-process peak \citep{1998ApJ...497..388G}. \citet{2013ApJ...774...98P} noted, using theoretical models of AGB stars, that rotation could decrease the final [Pb/Fe] abundance, down to a value of 1.6 (from 2.8 for the model with no rotation) for a 1.5~$\Msun$ model with a rotation velocity of 120~km~s$^{-1}$ and [Fe/H] of $-1.7$. The presence of rotation also decreases the [$hs$/$ls$] and [Pb/$hs$] ratios.

 The observed values for [C/Fe] are lower than the predictions of the best-matched model whereas [O/Fe] is observed to be overabundant. 
One possibility for the high [O/Fe] abundance is that the initial composition for the post-AGB stars was enhanced in O and Si compared to the scaled-solar initial composition used in the models. The required enhancements in the initial composition of [O/Fe] to match the abundances of the post-AGB stars range from 0.41 to 0.58. High abundances of other $\alpha$ elements (Mg, Ca, and Ti) are not observed in the post-AGB stars.

\section{Discussion of uncertainties}

There are many uncertainties in the input physics used for modelling AGB stars including convection, mass loss, extra mixing, reaction rates, rotation, and low-temperature opacities.
 It is therefore important to understand the role that these uncertainties have on the theoretical predictions.
 In this section we focus on a few uncertainties that can substantially affect the calculation of yields for neutron-capture elements; namely, mass loss, reaction rates, convection (and TDU), and the addition of a $^{13}$C pocket.

The mass-loss rate affects the lifetime of the AGB phase and the number of TDU episodes experienced. Therefore, mass loss plays a key role in influencing the chemical yields \citep[cf.][]{2007MNRAS.375.1280S,Karakas:2010et}. With a more efficient mass-loss rate, and hence a shorter AGB phase, lower yields of neutron-capture elements  are expected as  a smaller amount of enriched material is dredged to the surface to be expelled into the interstellar medium. Mass-loss rates are difficult to determine from observations and require the modelling of dust properties and radiative transfer. \citet{2009ApJ...696..797C} compared models of 2~$\Msun$ with $Z~=~0.0001$ using two different mass-loss prescriptions: one with a \citet[][$\eta=0.4$]{1975MSRSL...8..369R} prescription, and the standard case which uses a prescription similar to \citet{Vassiliadis:1993jk} but updated with more recent infrared observations \citep[see][]{2006NuPhA.777..311S}. The model with the \citet{1975MSRSL...8..369R} prescription shows an increase in the final surface abundance of the neutron-capture elements as a result of a longer AGB lifetime. However, it was found that the $s$-process indicators, [$hs$/$ls$] and [Pb/$hs$], are less sensitive to the duration of the AGB phase as the largest $^{13}$C pockets occur in the first few TPs and produce the largest increase in the $s$-process abundances. This sets the abundance ratio of the $s$-process indicators early in the AGB phase for the low-mass models. For the intermediate-mass models presented here, the values of the $s$-process indicators decrease over time and do not reach a constant ratio (see Figure~\ref{fig:f11}).

The uncertainties in the reaction rates can also have an impact on the production of the neutron-capture elements.
 In particular for $s$-process nucleosynthesis, the reaction rates of the neutron sources, $^{13}$C($\alpha$,$n$)$^{16}$O and $^{22}$Ne($\alpha$,$n$)$^{25}$Mg, can affect the number of neutrons produced per Fe seed.  The $^{13}$C($\alpha$,$n$)$^{16}$O reaction can only be measured experimentally at high energies and extrapolated to energies that occur during $s$-process nucleosynthesis. \citet{2012ApJ...756..193G} present an updated measurement for the reaction rate of $^{13}$C($\alpha$,$n$)$^{16}$O and compares to previous measurements using AGB models with $s$-process nucleosynthesis. Relative to the reaction rates from \citet{1988ADNDT..40..283C} and \citet{1999NuPhA.656....3A}, they find that if all the $^{13}$C is destroyed in radiative conditions minimal variations of up to 5 per cent occur for the neutron-capture elements. If some $^{13}$C is destroyed inside a convective TP, the updated reaction rate has a larger effect on the  abundance of the neutron-capture elements, with up to 25 per cent variation for Pb. The conditions where $^{13}$C is burnt convectively occur in low-mass stars  when there is incomplete radiative burning of $^{13}$C during an interpulse or if there is proton ingestion in a TP. In the models presented here, we use the $^{13}$C($\alpha$,$n$)$^{16}$O reaction rate taken from \citet{2008PhRvC..78b5803H} which is consistent with the updated measurement presented by \citet{2012ApJ...756..193G}. 

Concerning the uncertainties associated with convection and mixing length theory, a different $\alpha$ value will alter the amount of material mixed to the surface \citep[cf.][]{1988ApJ...328..671B}. \citet{2009ApJ...696..797C} investigated a 2~$\Msun$ model of $Z~=~0.0001$ with two different $\alpha$ values: 1.8 and 2.15. The lower $\alpha$ value resulted in lower temperatures at the bottom of the pulse driven convective zone produced by a TP. Another consequence of the lower $\alpha$ value was less efficient TDU with $9.54 \times 10^{-2}$~$\Msun$ dredged up compared to $1.6 \times 10^{-1}$~$\Msun$. This decrease in $M_{\rm TDU}$ occurs despite the standard model having one extra TP and results in lower final abundances. The $s$-process indicators for the low-mass models are less affected as they are more sensitive to the metallicity and to the mass of the $^{13}$C pocket.

As mentioned in Section~1, low-mass models require extra mixing of protons to form the $^{13}$C pocket \citep{Karakas:2014cn}.  We have shown in Section~\ref{sec:pmz} that changing the PMZ mass has an effect on the final surface abundances, particularly on the abundance of the neutron-capture elements. \citet{2013MNRAS.431.2861S} investigated varying the mass of the PMZ to match abundances in planetary nebulae and found that the predicted Ne abundance is sensitive to the size of the $^{13}$C pocket. The lack of understanding of the mechanism responsible for the formation of the $^{13}$C pocket highlights the uncertainties related to predicting yields of $s$-process nucleosynthesis. We refer the reader to \citet{2014ApJ...787...10B} and \citet{2014ApJ...787...41T} for further discussion on this point.  It is important to note that TDU and the formation of the $^{13}$C pocket should not be treated separately as is done with an added PMZ, particularly if the timescale for burning is shorter than the mixing timescale \citep{2004A&A...421L..25G}.

 The effect of rotation on the production of neutron-capture elements in AGB models has been studied by \citet{2003ApJ...593.1056H}, \citet{2004A&A...415.1089S}, and \citet{2013ApJ...774...98P}. It was determined that rotation reduces the neutron flux as the $^{13}$C pocket is mixed with the neutron poison $^{14}$N. This reduction in the number of neutrons in turn  may hinder the synthesis of the neutron-capture elements.  The presence of rotation offers a possible solution to the lower than predicted Pb abundances in post-AGB stars. We do not consider rotation in our models.

\section{Conclusions}

We have presented new AGB stellar models for a range of initial masses from 1~$\Msun$ to 7~$\Msun$ for a metallicity of $Z=0.001$  ([Fe/H] = $-1.2$) and a scaled-solar initial composition. In particular, $s$-process nucleosynthesis predictions for intermediate-mass AGB models of $Z~=~0.001$ are presented for the first time in the literature. We also present neutron-capture abundances and yields for a super-AGB model of 7~$\Msun$ for the first time. Online tables are available presenting (for each stellar mass) evolutionary properties, final surface abundances (including [X/H] and [X/Fe]) and yields for all elements, as well as isotope final surface abundances and yields for elements up to the Fe group.

We have presented in detail two representative AGB models, one low-mass model of 2~$\Msun$ and one intermediate-mass model of 5~$\Msun$. As a result of the activation of different neutron sources these models produce dissimilar abundance distributions. The low-mass models favour the production of Pb due to the $^{13}$C($\alpha$,$n$)$^{16}$O reaction whereas the intermediate-mass models favour the production of Rb over other neutron-capture elements due to the activation of branching points by the $^{22}$Ne($\alpha$,$n$)$^{25}$Mg  reaction. The [Rb/Zr] ratio, comparing two first $s$-process peak elements, is mass dependent with the intermediate-mass models showing an enhancement of $\sim 0.4$~dex. The low-mass models show a sub-solar value down to $-0.8$~dex for the 2~$\Msun$ model. 

The new predictions are compared to the $Z~=~0.001$ models of \citet{2009ApJ...696..797C,2011ApJS..197...17C} and \citet{2013MNRAS.434..488M}, for masses in common between the various studies. The differences in the final surface abundances and yields between the calculations can be attributed to the choice of input physics such as the treatment of convective borders. The elemental yield predictions of the models presented here are comparable to those by \citet{2011ApJS..197...17C}. The $s$-process indicators [$ls$/Fe] and [$hs$/Fe] agree to within 0.36 dex, with the largest difference occurring for the 1.5~$\Msun$ models for [$hs$/Fe]. For [Pb/Fe], the difference is less than 0.19~dex.

 We also investigated the uncertainty in the addition of a $^{13}$C pocket by varying the mass of the PMZ in the 3~$\Msun$ model. The 3~$\Msun$ model is in the transition zone between the lower mass models and the more massive models. Increases in the mass of the PMZ  result in enhancements in the abundances of neutron-capture elements and a number of light elements (O, F, Ne, Na, Mg, Al, and P). The intrinsic $s$-process indicator [$hs$/$ls$] is shown to be weakly dependent on the mass of the PMZ whereas [Pb/$hs$] decreases with increasing PMZ mass for the 3~$\Msun$ model due to lower neutron exposures.

One application of the AGB stellar models presented is a comparison of three low-metallicity post-AGB stars to the model predictions. Other applications include chemical evolution studies \citep[e.g.][]{2014ApJ...787...10B}  and the study of planetary nebulae in our Galaxy \citep[e.g.][]{2010PASA...27..227K} as well as external galaxies.  The models presented here have been used in the interpretation of measured abundances of globular cluster stars \citep{2014MNRAS.tmp..303Y,2014MNRAS.441.3396Y}, a chemically peculiar star in the Aquarius co-moving group \citep{Casey:2013uo}, and the $s$-process component of M4 and M22 \citep{shingles2014}.

\acknowledgments
\section*{Acknowledgments}

We thank the referee whose thorough comments have improved the clarity of the paper. The authors are grateful for the support of the NCI National Facility at the ANU. CKF is grateful for the financial support through an ANU PhD scholarship. AIK and ML are supported through an Australian Research Council Future Fellowship (FT110100475 and FT100100305, respectively). This research has made use of NASA's Astrophysics Data System.

\bibliographystyle{apj}

\appendix

 \begin{sidewaystable}
   \def\arraystretch{1.2}
   \caption{An example of the online data table providing the structural properties for each TP of the 1~$\Msun$ model.} 
 \label{tab:dataset1}
 \begin{center}
     \begin{tabular}{ccccccccccccccccc}
  \tableline\tableline
	$M_{\rm i}$$^{\rm a}$ &
	TP$^{\rm b}$ & 				% pulse number
	$M_{\rm core}$$^{\rm c}$ & 	% core mass
	$M_{\rm csh}$$^{\rm d}$ & % maximum mass of the intershell convection zone
	$\tau_{\rm csh}$$^{\rm e}$ & 				% duration of intershell convection
	$M_{\rm dredge}$$^{\rm f}$ &		% mass dredged into the envelope
	$\lambda$$^{\rm g}$ & 		% the TDU efficiency			
	$T_{\rm He}^{\rm max}$$^{\rm h}$ & 			% maximum temperature in the He-shell
	$T_{\rm TBCE}^{\rm max}$$^{\rm i}$ & 		% maximum temperature at the base of the convective envelope during the previous interpulse period
	$T_{\rm He}^{\rm max,ip}$$^{\rm j}$& 		% maximum temperature in the H-shell during the previous interpulse period
	$\tau_{\rm ip}$$^{\rm k}$	 & 		% interpulse period
	$M_{\rm tot}$$^{\rm l}$ & 		% total mass
	$L_{\rm max}$$^{\rm m}$ & 	%  maximum radiated luminosity during the interpulse period
	$L^{\rm max}_{\rm He}$$^{\rm n}$ & 		% maximum He-luminosity during a TP
	$R_{\rm max}$$^{\rm o}$ &		% maximum radius during the previous interpulse period
	$m_{\rm bol}$$^{\rm p}$ &		% bolometric magnitude
	$T_{\rm eff}$$^{\rm q}$  \\		% effective temperature at maximum radius
	(M$_{\odot}$) &
	 & 				% pulse number
	(M$_{\odot}$) & 	% core mass
	(M$_{\odot}$) & % maximum mass of the intershell convection zone
	(yr) & 				% duration of intershell convection
	(M$_{\odot}$) & 			% mass dredged into the envelope
	 & 		% the TDU efficiency			
	(K) & 			% maximum temperature in the He-shell
	(K) & 		% maximum temperature at the base of the convective envelope during the previous interpulse period
	(K) & 		% maximum temperature in the H-shell during the previous interpulse period
	(yr) & 		% interpulse period
	($M_{\odot}$) & 		% total mass
	(L$_{\odot}$) & 	%  maximum radiated luminosity during the interpulse period
	(L$_{\odot}$) & 		% maximum He-luminosity during a TP
	(R$_{\odot}$) &		% maximum radius during the previous interpulse period
	(mag.) &		% bolometric magnitude
	(K) \\		% effective temperature at maximum radius
\hline
1.0 & 1 & 5.29e$-$1 & 4.18e$-$2 & 5.22e+2 & 0.00e+0 & 0.00e+0 & 1.96e+8 & 1.18e+6 & 5.32e+7 & 0.00e+0 & 9.03e$-$1 & 1.58e+3 & 2.43e+5 & 7.98e+1 & $-$3.28e+0 & 4.15e+3   \\
1.0 & 2 & 5.32e$-$1 & 4.06e$-$2 & 4.18e+2 & 0.00e+0 & 0.00e+0 & 2.07e+8 & 1.21e+6 & 5.51e+7 & 2.39e+5 & 9.03e$-$1 & 1.86e+3 & 6.64e+5 & 8.94e+1 & $-$3.46e+0 & 4.09e+3   \\
1.0 & 3 & 5.39e$-$1 & 4.21e$-$2 & 5.30e+2 & 0.00e+0 & 0.00e+0 & 2.27e+8 & 1.30e+6 & 5.86e+7 & 3.17e+5 & 9.03e$-$1 & 2.55e+3 & 4.66e+6 & 1.11e+2 & $-$3.80e+0 & 3.98e+3   \\
1.0 & 4 & 5.47e$-$1 & 3.97e$-$2 & 4.80e+2 & 0.00e+0 & 0.00e+0 & 2.37e+8 & 1.34e+6 & 6.05e+7 & 3.32e+5 & 9.03e$-$1 & 2.99e+3 & 8.18e+6 & 1.24e+2 & $-$3.97e+0 & 3.92e+3   \\
1.0 & 5 & 5.57e$-$1 & 3.74e$-$2 & 4.85e+2 & 0.00e+0 & 0.00e+0 & 2.45e+8 & 1.38e+6 & 6.23e+7 & 3.12e+5 & 9.03e$-$1 & 3.45e+3 & 1.41e+7 & 1.38e+2 & $-$4.13e+0 & 3.88e+3   \\
1.0 & 6 & 5.67e$-$1 & 3.48e$-$2 & 3.97e+2 & 0.00e+0 & 0.00e+0 & 2.52e+8 & 1.41e+6 & 6.38e+7 & 2.84e+5 & 9.03e$-$1 & 3.91e+3 & 1.98e+7 & 1.50e+2 & $-$4.26e+0 & 3.84e+3   \\
1.0 & 7 & 5.77e$-$1 & 3.24e$-$2 & 3.86e+2 & 0.00e+0 & 0.00e+0 & 2.57e+8 & 1.43e+6 & 6.52e+7 & 2.53e+5 & 9.03e$-$1 & 4.36e+3 & 2.61e+7 & 1.62e+2 & $-$4.38e+0 & 3.80e+3   \\
1.0 & 8 & 5.88e$-$1 & 3.02e$-$2 & 3.44e+2 & 0.00e+0 & 0.00e+0 & 2.61e+8 & 1.43e+6 & 6.65e+7 & 2.25e+5 & 9.02e$-$1 & 4.81e+3 & 3.32e+7 & 1.74e+2 & $-$4.49e+0 & 3.77e+3   \\
1.0 & 9 & 5.98e$-$1 & 2.80e$-$2 & 2.95e+2 & 7.84e$-$4 & 7.56e$-$2 & 2.66e+8 & 1.42e+6 & 6.77e+7 & 1.99e+5 & 9.01e$-$1 & 5.27e+3 & 3.79e+7 & 1.85e+2 & $-$4.58e+0 & 3.75e+3   \\
1.0 & 10 & 6.07e$-$1 & 2.59e$-$2 & 2.52e+2 & 7.36e$-$4 & 7.30e$-$2 & 2.67e+8 & 1.33e+6 & 6.73e+7 & 1.66e+5 & 8.97e$-$1 & 5.99e+3 & 4.04e+7 & 2.10e+2 & $-$4.72e+0 & 3.63e+3   \\
1.0 & 11 & 6.16e$-$1 & 2.40e$-$2 & 1.98e+2 & 0.00e+0 & 0.00e+0 & 2.69e+8 & 1.14e+6 & 6.72e+7 & 1.43e+5 & 8.83e$-$1 & 6.59e+3 & 4.03e+7 & 2.44e+2 & $-$4.83e+0 & 3.50e+3   \\
1.0 & 12 & 6.25e$-$1 & 2.23e$-$2 & 2.04e+2 & 0.00e+0 & 0.00e+0 & 2.71e+8 & 1.01e+6 & 6.80e+7 & 1.27e+5 & 8.58e$-$1 & 6.99e+3 & 4.04e+7 & 2.55e+2 & $-$4.89e+0 & 3.48e+3   \\
1.0 & 13 & 6.34e$-$1 & 2.09e$-$2 & 1.87e+2 & 0.00e+0 & 0.00e+0 & 2.74e+8 & 8.48e+5 & 6.89e+7 & 1.15e+5 & 8.11e$-$1 & 7.40e+3 & 4.24e+7 & 2.65e+2 & $-$4.95e+0 & 3.48e+3   \\
1.0 & 14 & 6.43e$-$1 & 1.96e$-$2 & 1.74e+2 & 0.00e+0 & 0.00e+0 & 2.76e+8 & 6.32e+5 & 6.96e+7 & 1.05e+5 & 7.36e$-$1 & 7.79e+3 & 4.58e+7 & 2.63e+2 & $-$5.01e+0 & 3.53e+3   \\
1.0 & 15 & 6.51e$-$1 & 1.85e$-$2 & 1.62e+2 & 0.00e+0 & 0.00e+0 & 2.78e+8 & 3.88e+5 & 7.04e+7 & 9.59e+4 & 6.95e$-$1 & 8.17e+3 & 4.98e+7 & 2.39e+2 & $-$5.06e+0 & 3.69e+3   \\
1.0 & 16 & 6.59e$-$1 & 1.75e$-$2 & 1.56e+2 & 0.00e+0 & 0.00e+0 & 2.81e+8 & 2.59e+5 & 7.12e+7 & 8.78e+4 & 6.81e$-$1 & 8.56e+3 & 5.25e+7 & 2.16e+2 & $-$5.11e+0 & 3.90e+3   \\
1.0 & 17 & 6.67e$-$1 & 1.66e$-$2 & 1.53e+2 & 0.00e+0 & 0.00e+0 & 2.84e+8 & 2.14e+5 & 7.19e+7 & 8.02e+4 & 6.78e$-$1 & 8.96e+3 & 5.54e+7 & 1.87e+2 & $-$5.16e+0 & 4.30e+3   \\
\tableline
  \end{tabular} 
\\ NB. -- a) Initial stellar mass, b) pulse number, c) core mass, d) maximum mass of the intershell convection zone, e) duration of intershell convection, f) mass dredged into the envelope, g) the TDU efficiency, h) maximum temperature in the He-shell, i) maximum temperature at the base of the convective envelope during the previous interpulse period, j) maximum temperature in the H-shell during the previous interpulse period, k) interpulse period, l) total mass, m) maximum radiated luminosity during the previous interpulse period, n) maximum He-luminosity during a TP, o) maximum radius during the previous interpulse period, p) bolometric magnitude, and q) effective temperature at maximum radius. (Table~\ref{tab:dataset1} is published in its entirety in the electronic edition of ApJ. A portion is shown here for guidance regarding its form and content.)
 \end{center}
\end{sidewaystable}

 \begin{table*}
   \def\arraystretch{1.2}
 \begin{center}
  \caption{An example of the online data table providing the final surface abundances (in $Y$) for each isotope.}
 \label{tab:dataset2}
   \begin{tabular}{cccccccc}
  \tableline\tableline
Species & $A$ & 1~$\Msun$ & 1.25~$\Msun$ & 1.5~$\Msun$ & 2~$\Msun$ & 2.25~$\Msun$ & 2.5~$\Msun$ \\
\hline
p & 1 & 7.18e-1 & 7.09e-1 & 7.03e-1 & 6.81e-1 & 6.76e-1 & 6.85e-1  \\
d & 2 & 5.92e-7 & 5.59e-7 & 4.67e-7 & 1.03e-8 & 9.32e-10 & 1.07e-13  \\
$^3$He & 3 & 9.00e-5 & 8.87e-5 & 8.71e-5 & 6.26e-5 & 5.33e-5 & 5.01e-5  \\
$^4$He & 4 & 7.00e-2 & 7.15e-2 & 7.21e-2 & 7.55e-2 & 7.63e-2 & 7.44e-2  \\
$^7$Li & 7 & 2.89e-10 & 1.03e-10 & 6.62e-11 & 3.41e-11 & 2.00e-11 & 3.45e-12  \\
$^7$Be & 7 & 0.00e+0 & 0.00e+0 & 5.81e-31 & 0.00e+0 & 2.07e-38 & 7.41e-32  \\
$^8$B & 8 & 4.60e-26 & 2.89e-26 & 1.92e-27 & 7.72e-29 & 2.97e-29 & 3.52e-30  \\
$^{12}$C & 12 & 8.22e-5 & 2.62e-4 & 5.39e-4 & 1.09e-3 & 1.23e-3 & 1.15e-3  \\
$^{13}$C & 13 & 5.15e-7 & 4.55e-7 & 4.37e-7 & 4.07e-7 & 3.94e-7 & 4.19e-7  \\
$^{14}$C & 14 & 1.32e-20 & 2.03e-19 & 2.27e-13 & 2.21e-12 & 3.62e-12 & 1.72e-11  \\
$^{13}$N & 13 & 0.00e+0 & 0.00e+0 & 0.00e+0 & 0.00e+0 & 0.00e+0 & 0.00e+0  \\
$^{14}$N & 14 & 1.01e-5 & 1.03e-5 & 1.06e-5 & 1.17e-5 & 1.15e-5 & 1.04e-5  \\
$^{15}$N & 15 & 5.00e-9 & 6.38e-9 & 7.45e-9 & 5.89e-9 & 4.25e-9 & 3.26e-9  \\
\tableline
  \end{tabular} 
\\ NB.  -- Abundances of isotopes up to and including $^{70}$Zn are available online for all models. $Y = X/A$ where $X$ is the mass fraction and $A$ is the atomic mass.
(Table~\ref{tab:dataset2} is published in its entirety in the electronic edition of ApJ. A portion is shown here for guidance regarding its form and content.)
 \end{center}
\end{table*}

 \begin{table*}
   \def\arraystretch{1.2}
 \begin{center}
  \caption{An example of the online data table providing isotopic yields.}
 \label{tab:dataset3}
   \begin{tabular}{ccccccccc}
  \tableline\tableline
	Mass\tablenotemark{a} &
	Species\tablenotemark{b} & 
	$A$\tablenotemark{c} & 
	$M_i$\tablenotemark{d} & 
	$M_{\rm lost}(i)$\tablenotemark{e} & 
	$M_0 (i)$\tablenotemark{f} & 
	$\langle X(i) \rangle$\tablenotemark{g} & 
	$X_0(i)$\tablenotemark{h} & 			
	$f$\tablenotemark{i} \\
\hline
1.0 & n & 1 & 0.00e+0 & 0.00e+0 & 0.00e+0 & 0.00e+0 & 0.00e+0 & 0.00e+0   \\
1.0 & p & 1 & $-$8.78e$-$3 & 2.41e$-$1 & 2.49e$-$1 & 7.23e$-$1 & 7.49e$-$1 & $-$1.56e$-$2   \\
1.0 & d & 2 & $-$6.02e$-$7 & 4.82e$-$7 & 1.08e$-$6 & 1.45e$-$6 & 3.26e$-$6 & $-$3.52e$-$1   \\
1.0 & $^3$He & 3 & 8.90e$-$5 & 9.59e$-$5 & 6.89e$-$6 & 2.88e$-$4 & 2.07e$-$5 & 1.14e+0   \\
1.0 & $^4$He & 4 & 8.47e$-$3 & 9.17e$-$2 & 8.32e$-$2 & 2.75e$-$1 & 2.50e$-$1 & 4.21e$-$2   \\
1.0 & $^7$Li & 7 & 9.56e$-$11 & 2.87e$-$10 & 1.91e$-$10 & 8.63e$-$10 & 5.75e$-$10 & 1.76e$-$1   \\
1.0 & $^7$Be & 7 & 7.94e$-$12 & 7.94e$-$12 & 0.00e+0 & 2.38e$-$11 & 0.00e+0 & 0.00e+0   \\
1.0 & $^8$B & 8 & 9.20e$-$26 & 9.20e$-$26 & 0.00e+0 & 2.76e$-$25 & 0.00e+0 & 0.00e+0   \\
1.0 & $^{12}$C & 12 & 1.90e$-$4 & 2.47e$-$4 & 5.70e$-$5 & 7.43e$-$4 & 1.71e$-$4 & 6.38e$-$1   \\
1.0 & $^{13}$C & 13 & 1.32e$-$6 & 2.01e$-$6 & 6.90e$-$7 & 6.04e$-$6 & 2.07e$-$6 & 4.65e$-$1   \\
1.0 & $^{14}$C & 14 & 1.39e$-$9 & 1.39e$-$9 & 0.00e+0 & 4.16e$-$9 & 0.00e+0 & 0.00e+0   \\
1.0 & $^{13}$N & 13 & 0.00e+0 & 0.00e+0 & 0.00e+0 & 0.00e+0 & 0.00e+0 & 0.00e+0   \\
1.0 & $^{14}$N & 14 & 1.95e$-$5 & 3.64e$-$5 & 1.68e$-$5 & 1.09e$-$4 & 5.06e$-$5 & 3.35e$-$1   \\
1.0 & $^{15}$N & 15 & $-$1.41e$-$8 & 2.63e$-$8 & 4.03e$-$8 & 7.89e$-$8 & 1.21e$-$7 & $-$1.86e$-$1   \\
\tableline
  \end{tabular} 
\\ NB. -- Yields of isotopes up to and including $^{70}$Zn are available online for all models. a) Initial stellar mass, b) species $i$, c) mass number, d) net stellar yield as defined in Equation (4), e) amount of the species $i$ in the wind lost from the star, f) total mass expelled during the stellar lifetime multiplied by the initial mass fraction, g) average mass fraction of species $i$ in the wind, h) initial mass fraction of species $i$, i) production factor $f$ defined as $\log_{10}[\langle X(i)\rangle/X_0(i)]$.
 (Table~\ref{tab:dataset3} is published in its entirety in the electronic edition of ApJ. A portion is shown here for guidance regarding its form and content.)
 \end{center}
\end{table*}

 \begin{table*}
   \def\arraystretch{1.2}
 \begin{center}
  \caption{An example of the online data table providing the final surface abundance of each element.}
 \label{tab:dataset4}
   \begin{tabular}{cccccccc}
  \tableline\tableline
\colhead{Mass\tablenotemark{a}} &
	El.\tablenotemark{b} & 
	$Z$\tablenotemark{c} & 
	$\log~\epsilon(X)$\tablenotemark{d} & 
	[X/H]\tablenotemark{e} & 
	[X/Fe]\tablenotemark{e} & 
	[X/O]\tablenotemark{e} & 			
	$X(i)$\tablenotemark{h} \\
\hline
1.0 & H & 1 & 1.20e+1 & 0.00e+0 & 0.00e+0 & 0.00e+0 & 7.24e$-$1  \\
1.0 & He & 2 & 1.10e+1 & 5.97e$-$2 & 1.23e+0 & 1.21e+0 & 2.81e$-$1  \\
1.0 & Li & 3 & 2.60e+0 & $-$6.56e$-$1 & 5.11e$-$1 & 4.99e$-$1 & 2.00e$-$9  \\
1.0 & Be & 4 & 0.00e+0 & $-$1.30e+0 & 0.00e+0 & 0.00e+0 & 0.00e+0  \\
1.0 & B & 5 & 0.00e+0 & $-$2.79e+0 & 0.00e+0 & 0.00e+0 & 4.97e$-$25  \\
1.0 & C & 6 & 8.06e+0 & $-$4.08e$-$1 & 7.59e$-$1 & 7.46e$-$1 & 9.93e$-$4  \\
1.0 & N & 7 & 7.15e+0 & $-$7.23e$-$1 & 4.44e$-$1 & 4.31e$-$1 & 1.41e$-$4  \\
1.0 & O & 8 & 7.58e+0 & $-$1.15e+0 & 1.27e$-$2 & 0.00e+0 & 4.32e$-$4  \\
1.0 & F & 9 & 3.55e+0 & $-$8.71e$-$1 & 2.96e$-$1 & 2.83e$-$1 & 4.82e$-$8 \\
1.0 & Ne & 10 & 6.89e+0 & $-$1.08e+0 & 8.51e$-$2 & 7.24e$-$2 & 1.12e$-$4  \\
1.0 & Na & 11 & 5.21e+0 & $-$1.03e+0 & 1.35e$-$1 & 1.22e$-$1 & 2.66e$-$6  \\
1.0 & Mg & 12 & 6.47e+0 & $-$1.17e+0 & $-$1.16e$-$3 & $-$1.39e$-$2 & 5.17e$-$5  \\
\tableline
  \end{tabular} 
\\ NB. -- Final surface abundances of elements up to and including Po are available online for all models.
a) Initial stellar mass, b) element, c) atomic number, d) $\log~\epsilon(X) = \log_{10}(N_A/N_H) + 12$ where $N_A$ and $N_H$ are abundances of element $A$ and H, e) [X/Y] = $\log_{10}(N_X/N_Y)_{\star} - \log_{10}(N_X/N_Y)_{\odot}$ where $N_X$ and $N_Y$ are the abundances of elements $X$ and $Y$, and f) mass fraction of element.
 (Table~\ref{tab:dataset4} is published in its entirety in the electronic edition of ApJ. A portion is shown here for guidance regarding its form and content.)
 \end{center}
\end{table*}

 \begin{table*}
   \def\arraystretch{1.2}
 \begin{center}
  \caption{An example of the online data table providing elemental yields.}
 \label{tab:dataset5}
   \begin{tabular}{ccccccccc}
  \tableline\tableline
	Mass\tablenotemark{a} &
	Species\tablenotemark{b} & 
	$A$\tablenotemark{c} & 
	$M_i$\tablenotemark{d} & 
	$M_{\rm lost}(i)$\tablenotemark{e} & 
	$M_0 (i)$\tablenotemark{f} & 
	$\langle X(i) \rangle$\tablenotemark{g} & 
	$X_0(i)$\tablenotemark{h} & 			
	$f$\tablenotemark{i} \\
\hline
1.0 & n & 0 & 0.00e+0 & 0.00e+0 & 0.00e+0 & 0.00e+0 & 0.00e+0 & 0.00e+0 \\
1.0 & H & 1 & $-$8.78e$-$3 & 2.41e$-$1 & 2.49e$-$1 & 7.23e$-$1 & 7.49e$-$1 & $-$1.56e$-$2 \\
1.0 & He & 2 & 8.56e$-$3 & 9.18e$-$2 & 8.32e$-$2 & 2.76e$-$1 & 2.50e$-$1 & 4.25e$-$2 \\
1.0 & Li & 3 & 9.56e$-$11 & 2.87e$-$10 & 1.91e$-$10 & 8.63e$-$10 & 5.75e$-$10 & 1.76e$-$1 \\
1.0 & C & 6 & 1.92e$-$4 & 2.49e$-$4 & 5.77e$-$5 & 7.49e$-$4 & 1.73e$-$4 & 6.36e$-$1 \\
1.0 & N & 7 & 1.95e$-$5 & 3.64e$-$5 & 1.69e$-$5 & 1.09e$-$4 & 5.07e$-$5 & 3.34e$-$1 \\
1.0 & O & 8 & 3.59e$-$6 & 1.43e$-$4 & 1.40e$-$4 & 4.31e$-$4 & 4.20e$-$4 & 1.10e$-$2 \\
1.0 & F & 9 & 5.63e$-$9 & 1.38e$-$8 & 8.13e$-$9 & 4.13e$-$8 & 2.44e$-$8 & 2.29e$-$1 \\
1.0 & Ne & 10 & 5.05e$-$6 & 3.56e$-$5 & 3.06e$-$5 & 1.07e$-$4 & 9.18e$-$5 & 6.64e$-$2 \\
1.0 & Na & 11 & 1.37e$-$7 & 7.87e$-$7 & 6.50e$-$7 & 2.36e$-$6 & 1.95e$-$6 & 8.29e$-$2 \\
1.0 & Mg & 12 & $-$2.17e$-$8 & 1.72e$-$5 & 1.73e$-$5 & 5.18e$-$5 & 5.19e$-$5 & $-$5.45e$-$4 \\
\tableline
  \end{tabular} 
\\ NB. -- Yields of elements up to and including Po are available online for all models. a) Initial stellar mass, b) species $i$, c) mass number, d) net stellar yield as defined in Equation (4), e) amount of the species $i$ in the wind lost from the star, f) total mass expelled during the stellar lifetime multiplied by the initial mass fraction, g) average mass fraction of species $i$ in the wind, h) initial mass fraction of species $i$, i) production factor $f$ defined as $\log_{10}[\langle X(i)\rangle/X_0(i)]$.
 (Table~\ref{tab:dataset5} is published in its entirety in the electronic edition of ApJ. A portion is shown here for guidance regarding its form and content.)
 \end{center}
\end{table*}

\end{document}